\documentclass[journal,final]{IEEEtran}
\usepackage{cite}

\usepackage{marginnote}

\ifCLASSINFOpdf
   \usepackage[pdftex]{graphicx}
  \graphicspath{{./Images/Experiment/Rain_new/}{./Images/Experiment/Chimney_new/}{./Images/Simulation/New_four/}{./Images/Simulation/imag_20211108/} }
\else
   \usepackage[dvips]{graphicx}
\fi
\usepackage{marginnote}
\usepackage{xcolor}
\usepackage{booktabs}

\usepackage{amsmath}
\usepackage{amssymb}

\usepackage{mathrsfs}

\usepackage{multirow}

\usepackage[ruled,linesnumbered]{algorithm2e}

\usepackage{array}

\ifCLASSOPTIONcompsoc
  \usepackage[caption=false,font=normalsize,labelfont=sf,textfont=sf]{subfig}
\else
  \usepackage[caption=false,font=footnotesize]{subfig}
\fi

\usepackage{stfloats}
\begin{document}
%
\title{Interference Mitigation for FMCW Radar With Sparse and Low-Rank
	Hankel Matrix Decomposition}
%
%
%

\author{Jianping~Wang,~\IEEEmembership{Member, IEEE},
        Min~Ding,
        and~Alexander Yarovoy,~\IEEEmembership{Fellow,~IEEE}
\thanks{J. Wang and A. Yarovoy are with the Faculty of Electrical Engineering, Mathematics and Computer Science, Delft University of Technology, Delft, 2628CD, The Netherlands. e-mail: J.Wang-4@tudelft.nl, A.Yarovoy@tudelft.nl.}
\thanks{M. Ding was with the Faculty of Electrical Engineering, Mathematics and Computer Science, Delft University of Technology, Delft, 2628CD, The Netherlands, and now is with State Grid Shanghai Municipal Electric Power Company, Shanghai, 200122, China. email: min.dingchina@hotmail.com.}
}

\maketitle

\begin{abstract}
In this paper, the interference mitigation for Frequency Modulated Continuous Wave (FMCW) radar system with a dechirping receiver is investigated.  After dechirping operation, the scattered signals from targets result in beat signals, i.e., the sum of complex exponentials while the interferences lead to chirp-like short pulses. Taking advantage of these different time and frequency features between the useful signals and the interferences, the interference mitigation is formulated as an optimization problem: a sparse and low-rank decomposition of a Hankel matrix constructed by lifting the measurements. Then, an iterative optimization algorithm is proposed to tackle it by exploiting the Alternating Direction of Multipliers (ADMM) scheme. Compared to the existing methods, the proposed approach does not need to detect the interference and also improves the estimation accuracy of the separated useful signals. Both numerical simulations with point-like targets and experiment results with distributed targets (i.e., raindrops) are presented to demonstrate and verify its performance. The results show that the proposed approach is generally applicable for interference mitigation in both stationary and moving target scenarios.   
\end{abstract}


%
\IEEEpeerreviewmaketitle

\section{Introduction}
%
%
%
%
\IEEEPARstart{W}{ide} spread of Frequency Modulated Continuous Wave (FMCW) radar (automotive radar, vital sign observation, intelligent light, etc.) has raised serious concerns about their mutual interference. It was shown that in FMCW radar external interference increases overall clutter level, might result in appearance of ghost targets, masks weak targets and disturbs target returns. Numerous approaches have been proposed to mitigate the interference. However, desired level of interference suppression at given signal processing costs has not been achieved.

In the literature, the interference mitigation (IM) of the FMCW radar systems is generally tackled through two ways: (1) system/signal (i.e., antenna array, waveform, etc) design methods \cite{Xu2018GRS,Zhang2019IET,Uysal2020TVT,Kitsukawa2019EuRAD,Artyukhin2019EnT,Aydogdu2019ITS,Ishikawa2019RADAR,Irazoqui2019IEEEAcess,Kirk2017SPIE,Solodky2021} and (2) signal processing approaches \cite{Chen2018GRS,Nozawa2017Radar,Jin2019ANC,Babur2010,Neemat2019,Choi2016,JWang2020,Uysal2019,Ren2019,Nguyen2016,Su2017JSTARS,Li2015IET,Huang2019,Lee2019TITS}. The system/signal design related methods generally require that the radar system have the capability for agile waveform modulation, new radar architecture, spectrum sensing module or a coordination unit, etc.; thus, the radar system has to be modified or redesigned and the system complexity will increase. On the other hand, the signal processing approaches tackle the interference mitigation problem through post-processing, which could be readily used for the existing FMCW radar systems.
The work presented in this paper falls into the latter group. The signal processing approaches to interference mitigation of radar system can be categorized into three classes: filtering approaches \cite{Chen2018GRS,Nozawa2017Radar,Jin2019ANC,Liu2020Access}, interference nulling and reconstruction methods \cite{Babur2010,Neemat2019,Choi2016,JWang2020}, and signal separation approaches \cite{Uysal2019,Ren2019,Nguyen2016,Su2017JSTARS,Li2015IET,Huang2019,Lee2019TITS}. The filtering approaches usually exploit different distributions of interferences and useful signals in a specific domain (i.e., time, frequency or space) and design a proper filter to suppress the interferences. They work well for some fixed or slowly variant interferences. For complex scenarios, the interference could be transient or change rapidly between different sweeps. Then adaptive filtering could be utilized \cite{Jin2019ANC}, which uses a reference input to generate the correlated interference component for interference mitigation. However, it is not easy to find a proper reference input signal in practice; thus, its performance of interference mitigation is of no guarantee.

The interference nulling and reconstruction methods is to first cut out the contaminated signal samples \cite{Babur2010} and then reconstruct the signals in the cut-out region \cite{Neemat2019,JWang2020}. To these methods, precise detection of the interference and accurate reconstruction of the cut-out signal samples are the key steps. As the amplitudes of  interferences and the targets' signals in the low-pass filter output of the FMCW radar systems are generally unknown, it makes the detection of the interference a challenging problem. If the interferences cannot be precisely detected, they will be removed partially, or the useful signals are excessively eliminated \cite{Choi2016}. Moreover, even if the interferences are detected properly, the targets' signals contaminated by the interferences are unavoidably to be cut out, resulting in the power loss of the targets' signals. To fix this problem, model-based methods are proposed to reconstruct the useful signals in the cut-out region by using the Burg method \cite{Neemat2019} and Matrix pencil approach \cite{JWang2020}. However, with the increase of the proportion of the contaminated samples in the full measurements, the accuracy of the reconstructed signals with these approaches would decrease significantly .


On the other hand, signal separation approaches tackle the interference mitigation by exploiting the sparsity of the interferences or the useful signals in some bases (or domains), which circumvents explicit detection of interferences. For example, for ultra-wideband radar systems the useful signals of targets could exhibit as sparse spikes in time while the Radio Frequency Interferences (RFI) are generally sparse in the frequency domain \cite{Ren2019}. For FMCW radars, the beat signals could be sparse in frequency domain. By contrast, the interferences after dechirping and anti-aliasing low-pass filtering operations generally exhibit as chirp-like signals which could be sparse in the time-, frequency- or time-frequency domain in various scenarios. To exploit the sparse representations of beat signals in the frequency domain and of the interferences in the time-frequency domain, the Discrete Fourier Transform (DFT) bases and the Short-Time Fourier Transform (STFT) bases, which are regular grids in the corresponding domains, are generally employed and the Split Augmented Lagrangian Shrinkage Algorithm
(SALSA) could be used to solve the related morphological component analysis problem \cite{Uysal2019}. Although these representations are attractive for efficient implementation of the SALSA-based interference mitigation method by taking advantage of the Fast Fourier Transform (FFT), the inherent so-called ``off-grid'' problem, i.e., the mismatch between true frequency (or time-frequency) components with the discrete bases, could lead to less sparse representation and consequently degrade the interference mitigation performance.

Recently, transient interference or RFI suppression has also been tackled by formulating as a Robust Principle Component Analysis (RPCA) problem \cite{Candes2011,Ding2011TIP}, where a matrix constructed by the measurements is decomposed as the sum of a sparse matrix and a low-rank one \cite{Li2015IET,Nguyen2016,Su2017JSTARS,Huang2019}. The formulated RPCA problems are solved by using the singular valued thresholding (SVT) algorithm \cite{Nguyen2016,Li2015IET}, reweighted nuclear norm or reweighted Frobenius norm \cite{Huang2019}. In these approaches, the operations are directly involved in the singular values of the matrix constructed with the signals.
So the singular value decomposition (SVD) is required in these approaches, which is generally very computationally expensive, especially for large matrices. Moreover, as the signal matrix formed by the measurements in multiple pulse repetition intervals (PRI) for transient interference mitigation for synthetic aperture radar systems are typically not structured, the developed approaches scarcely exploit the structures of the matrices. 

{
To circumvent the ``off-grid'' problem of traditional signal separation methods and heavy computational load of the existing RPCA-based approaches, we propose a novel Interference Mitigation approach for FMCW radars with SPArse and low-Rank HanKeL matrix dEcomposition (IM-SPARKLE). For the proposed approach, we formulate the interference mitigation for FMCW radars as a RPCA-like problem by exploiting the time sparsity of interferences and spectral sparsity of useful signals.
Inspired by the matrix pencil approach for exponential component estimation \cite{Hua1990,JWang2020}, the spectral sparsity of the useful signal is exploited by minimizing the rank of a Hankel matrix constructed with its samples in the time domain. The rank minimization problem is generally relaxed as a nuclear norm minimization problem. To circumvent the computationally expensive SVD used for the nuclear norm minimization, we utilized a Frobenius norm factorization \cite{Jin2018,Srebro2004,Signoretto2013} to relax the nuclear norm minimization; thus, much more efficient algorithm is developed compared to the traditional RPCA-based IM approaches. In addition, our formulation directly imposes the sparsiy of interferences on its time-domain samples and circumvents the nonuniform weighting effect on different samples when a Hankel matrix of interference components is used. Thus, naturally there is no need to use the reweighting operations as in \cite{Huang2019}.   

 
}

{

The rest of the paper is organized as follows. In section~\ref{sec:problem_formulation}, the problem formulation of the interference mitigation as a signal separation problem is presented, which results in an optimization problem of a sparse and low rank decomposition of a Hankel matrix. Then, the proposed algorithm to solve this problem is provided in detail in section~\ref{sec:Approach}. After that, section \ref{sec:Simulation_results} and section \ref{sec:Experiments} show both numerical results and experimental measurements to demonstrate the performance of the proposed IM approach for FMCW radar systems. Finally, conclusions are drawn in section~\ref{sec:conclusion}.

}

\section{Problem formulation} \label{sec:problem_formulation}
For an interference-contaminated FMCW radar system, target responses are received together with interferences. The signal acquired with a deramping receiver can be expressed as
\begin{equation}  \label{eq:beatSig_continuous}
    y(t) = x(t) + i(t)+ n(t), 
\end{equation}
where $x(t) =  \sum_{i=1}^{M} \sigma_i \exp\left( -j2\pi f_{b,i}t \right)$ is the beat signal of targets with complex amplitudes $\sigma_i$ and beat frequencies $f_{b,i}$ after deramping operation. $i(t)$ is the interference, which generally has a short duration after deramping and low-pass filtering operations and exhibits as a sum of chirp-like signals \cite{JWang2020}. $n(t)$ represents the additive white complex Gaussian noise (AWGN) and measurement errors. For the discrete signals sampled with intervals $\Delta t$, \eqref{eq:beatSig_continuous} can be rewritten as
\begin{equation} \label{eq:beatSig_discrete}
y[k] = x[k] +i[k] + n[k]
\end{equation}    
where $k=0,1,\cdots,N-1$ denotes the indices of the discrete-time samples. Stacking all the measurements together, one can get
\begin{equation} \label{eq:beatSig_vector}
\mathbf{y} = \mathbf{x} + \mathbf{i} +\mathbf{n}
\end{equation}
where 
\begin{align*}
\mathbf{x} &= [x[0], x[1], \cdots, x[N-1]]^T \ \in \mathbb{C}^{N}, \\
\mathbf{i} &= [i[0], i[1], \cdots, i[N-1]]^T \ \in \mathbb{C}^{N},\\
\mathbf{n} &= [n[0], n[1], \cdots, n[N-1]]^T \ \in \mathbb{C}^{N},\\
\mathbf{y} &= [y[0], y[1],\cdots, y[N-1]]^T \ \in \mathbb{C}^{N}.
\end{align*} 

Considering the different time-frequency properties of the ``beat signals" resulting from targets' responses and interferences, the targets' beat signals are sparse in the frequency domain while the counterparts of interferences are sparse in the time-frequency domain, especially for point-like targets scenarios. Taking advantage of this feature, Uysal \cite{Uysal2019} has formulated the interference mitigation as a signal separation problem by pursuing sparse representations of targets' beat signals and interferences in two different sets of bases. It can be explicitly expressed as an optimization problem 
\begin{align} \label{eq:optimProblem_SALSAR}
\{\mathbf{a}_x, \mathbf{a}_i\} = \arg\min_{\mathbf{a}_x, \mathbf{a}_i} & \|\mathbf{a}_x\|_1 + \lambda \|\mathbf{a}_i\|_1 \\
\qquad \text{s.t.} \quad &\| \mathbf{y} - \mathbf{x} - \mathbf{i} \|_2^2 < \epsilon \nonumber \\
&\mathbf{x} = \mathbf{F}_x \mathbf{a}_x, \  \mathbf{i} = \mathbf{F}_i \mathbf{a}_i \nonumber 
\end{align} 
where $\mathbf{F}_x$ and $\mathbf{F}_i$ are the bases for the sparse representations of targets' beat signals and interferences, respectively.  $\mathbf{a}_x$ and $\mathbf{a}_i$ are the corresponding coefficients. $\lambda$ is the regularization parameter used to trade off between the two terms. Generally, the discrete Fourier transform basis is used as $\mathbf{F}_x$ and the Short-Time Fourier Transform (STFT) basis is employed as $\mathbf{F}_i$ \cite{Uysal2019}. Then, the optimization problem in \eqref{eq:optimProblem_SALSAR} can be efficiently addressed by incorporating the fast Fourier transform (FFT). However, due to the possible grid-off problem related to the FFT, the spectrum of a target's beat signal could spread over multiple Fourier grids. Thus, part of the signal spectrum may be erroneously decomposed to be interferences, which causes signal power low and degrades the interference suppression performance.    

\begin{figure}[!t]
	\centering
	\includegraphics[width=0.48\textwidth]{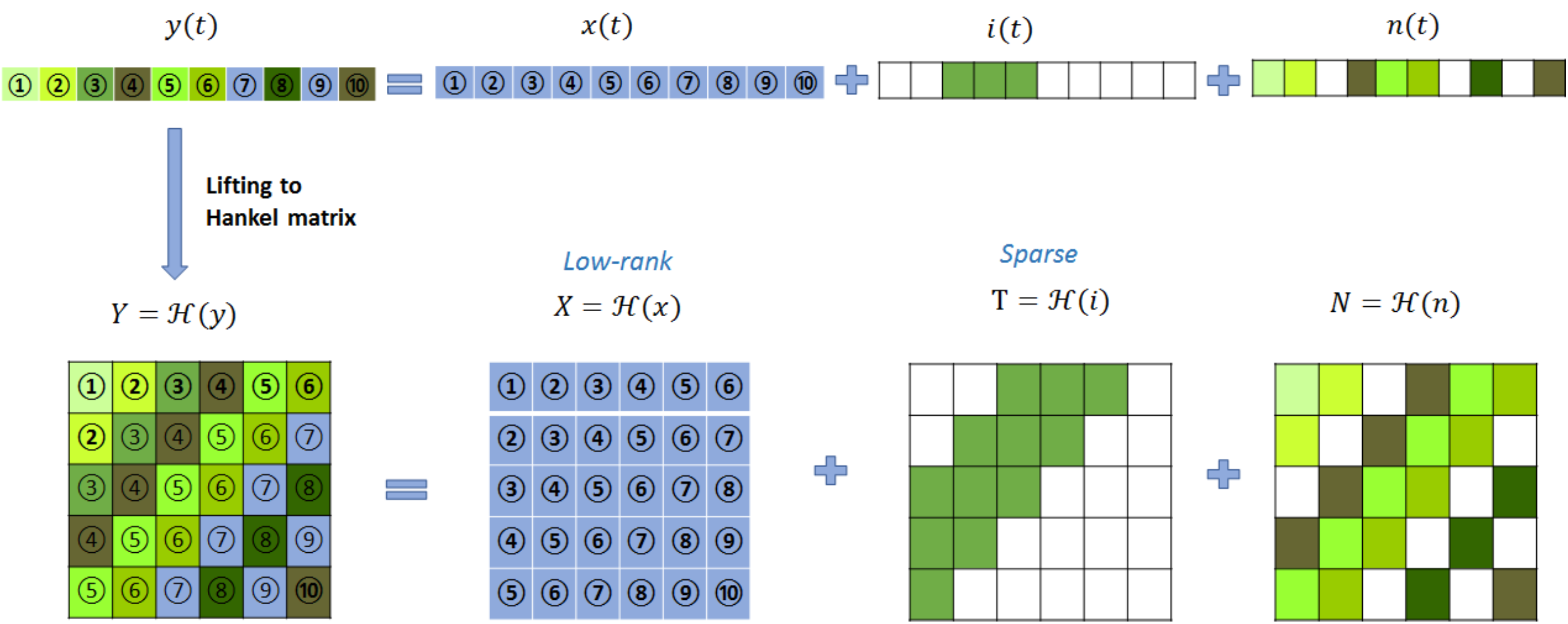}

	\caption{\label{fig:Graphic illustration}Graphic illustration of the sparse and low-rank decomposition of a  Hankel structured matrix from the interference-contaminated FMCW radar system.}
\end{figure}

The matrix pencil method \cite{Hua1990,Sarkar1995,Wang2018}  is an accurate approach for parameter estimations of complex exponentials by lifting the measurements as a Hankel matrix. 

Inspired by this idea, we lift the interference-contaminated measurements as a Hankel matrix (see Fig.~\ref{fig:Graphic illustration}). For the measurement vector $\mathbf{y}\in \mathbb{C}^{N\times 1}$, a Hankel matrix $\mathbf{Y}\in \mathbb{C}^{m\times n}$ can be constructed, where $N=m+n-1$, $m>M$ and $n>M$. According to \eqref{eq:beatSig_vector}, the Hankel matrix $\mathbf{Y}$ can be explicitly expressed as
\begin{align} \label{eq:beatSig_liftHankel}
\mathbf{Y} = \mathscr{H}(\mathbf{y})&=\mathscr{H}(\mathbf{x}) + \mathscr{H}(\mathbf{i}) +\mathscr{H}(\mathbf{n}) \nonumber \\
&=  \mathbf{X} + \mathbf{T} + \mathbf{N}
\end{align}
where $\mathscr{H}(\cdot)$ is the operator that lifts a vector as a Hankel matrix. $\mathbf{X} = \mathscr{H}(\mathbf{x})$, $\mathbf{T} = \mathscr{H}(\mathbf{i})$ and $\mathbf{N} = \mathscr{H}(\mathbf{n})$ are the Hankel matrices formed by the useful beat signals, interference and noise components, respectively.  The Hankel matrix $\mathbf{X}$ related to targets' beat signals is generally a low-rank matrix. Its rank is determined by the number of complex exponentials, i.e., beat signals in our case, which is usually much smaller than the matrix dimension. Meanwhile, the interference, as mentioned above, has a short duration in time; thus it is considered to be sparse in the time domain. As the construction of a Hankel matrix from a vector is a linear operation,  the time-sparse interference leads to a sparse Hankel matrix $\mathbf{T}$. This important observation motivates us to exploit the sparse and low-rank decomposition of a Hankel structured matrix to separate the useful signal and the interference from the measurement data. It can be expressed as
\begin{align} \label{eq:sigSep_matOpt_rank_L0}
\{\mathbf{X, T}\}=\arg\min_{\mathbf{X,T}} \ \ &\text{rank}\left(\mathbf{X}\right) + \tau \|\mathbf{T}\|_0 \\
\text{s.t.}\ \ & \|\mathbf{Y-X-T}\|_F^2 \leq \varepsilon  \nonumber \\
& \mathbf{X, T} \in \mathcal{H} \nonumber  
\end{align} 
where $\mathrm{rank}(\mathbf{X})=\sum_t |\sigma_t|_0$ is the number of non-zero singular values (i.e., rank) of the matrix $\mathbf{X}$, and $\|\mathbf{T}\|_0=\sum_{p=1}^m\sum_{q=1}^n |T_{pq}|_0$ (with slight abuse of notation) denotes the number of non-zero entries $T_{pq}$ in the matrix $\mathbf{T}$. 

As it is difficult to solve the minimization of the rank and $\ell_0$-norm sparse problem directly, 
the rank operation and $\ell_0$-norm are generally replaced, respectively, by the nuclear norm and $\ell_1$-norm for relaxation. Then, \eqref{eq:sigSep_matOpt_rank_L0} can be relaxed as 
\begin{alignat}{2} \label{eq:sigSep_matOpt_nuNorm_L1}
\left\{\mathbf{X,T}\right\}=\arg \min_{\mathbf{X,T}} \quad &\mathbf{\left\|X\right\|}_* +  \tau  \left\|\mathbf{T}\right\|_1 \\ 
\mbox{s.t.} \quad
& \left\|\mathbf{Y-X-T}\right\|_F^2  \leq \varepsilon \nonumber\\
&\mathbf{X, T} \in \mathcal{H} \nonumber
\end{alignat} 
where $\|\mathbf{X}\|_\ast = \sum_t|\sigma_t|_1$ denotes the nuclear norm, i.e., the sum of the singular values of the matrix $\mathbf{X}$, and $\|\mathbf{T}\|_1 = \sum_{p=1}^m\sum_{q=1}^n |T_{pq}|_1$ denotes the sum of absolute values (or, magnitudes) of the non-zeros entries of the matrix $\mathbf{T}$.

One can note that the formulation in \eqref{eq:sigSep_matOpt_nuNorm_L1} is a typical RPCA problem \cite{Candes2011}. So, the Hankel matrices $\mathbf{X}$ and $\mathbf{T}$ can be readily obtained by using the RPCA algorithm and then the separated beat signals and interferences can be extracted from their first columns and last rows, respectively. However, due to the different numbers of recurrence of the elements in the lifted Hankel matrix $\mathbf{T}$, the interference samples located close to the main anti-diagonal are implicitly weighted by a larger factor when the sparsity constraint is imposed. To circumvent this problem, we advise not lifting the interference part, i.e., replacing the $\ell_1$-norm of the lifted matrix $\mathbf{T}$ by the $\ell_1$-norm of the interference vector $\mathbf{i}$. Meanwhile, the Hankel matrix $\mathbf{X}$ can be substituted by $\mathscr{H}(\mathbf{x})$ to explicitly accounting for the Hankel structure constraint. As a result, \eqref{eq:sigSep_matOpt_nuNorm_L1} can be rewritten (with slight abuse of the notation $\varepsilon$) as 
\begin{align} \label{eq:sigSep_vecOpt_x_i}
\left\{\mathbf{x,i} \right\} = \arg \min_{\mathbf{x,i}} \quad &\left\|\mathscr{H}(\mathbf{x})\right\|_* +  \tau  \left\|\mathbf{i}\right\|_1\\
\text{s.t.} \quad
& \left\|\mathbf{y-x-i}\right\|_2^2  \leq \varepsilon \nonumber 
\end{align}    
The above optimization problem generally can be tackled by using the RPCA algorithm with some minor modifications. However, singular value decomposition is involved in the RPCA approach, which is very computationally expensive, especially for larger matrices.

Considering the complex exponential model of beat signals $x(t)$ in \eqref{eq:beatSig_continuous}, the Hankel matrix $\mathscr{H}(\mathbf{x})$ can be decomposed as
\begin{equation} \label{eq:HankelMat_VandeDecomposition}
    \mathscr{H}(\mathbf{x}) =  \mathbf{Z_L \Sigma_M Z_R}, 
\end{equation}
where 
\begin{equation}
    \mathbf{Z_L} =  
\left[\begin{array}{cccc} 
    1 &    1    & \cdots &   1\\ 
    z_1 &  z_2    & \cdots&    z_M\\ 
    \vdots  & \vdots  &       & \vdots \\
    z_1^{N-L-1}  & z_2^{N-L-1} & \cdots& z_M^{N-L-1}
\end{array}\right] ,
\end{equation}
\begin{equation}
    \mathbf{\Sigma_M} = \text{diag}\{\sigma_1, \sigma_2, \cdots, \sigma_M\},
\end{equation}
\begin{equation}
    \mathbf{Z_R} =  
\left[\begin{array}{cccc} 
   1 &    z_1    & \cdots &   z_1^{L-1}\\ 
    1 &  z_2    & \cdots&    z_2^{L-1}\\ 
    \vdots  & \vdots  &       & \vdots \\
    1  & z_M & \cdots& z_M^{L-1}
\end{array}\right] .
\end{equation}
and $z_i = \exp(-j2\pi f_{b,i} \Delta t)$ is the signal poles with respect to the corresponding target. Note that $\mathbf{Z_L} \in \mathbb{C}^{m\times M}$ and $\mathbf{Z_R} \in \mathbb{C}^{M\times n}$ are two Vandermonde matrices formed with the signal poles. If we decompose $\mathbf{\Sigma_M}$ as the square of a diagonal matrix $\mathbf{\bar{\Sigma}}$ (i.e., $\mathbf{\Sigma_M} = \mathbf{\bar{\Sigma}}^2$) and denote $\mathbf{U}=\mathbf{Z_L \bar{\Sigma}}$ and $\mathbf{V} = \left(\mathbf{\bar{\Sigma} Z_R} \right)^H$, then we have
\begin{equation} \label{eq:Hankel_X_UV}
\mathscr{H}(\mathbf{x}) = \mathbf{Z_L \bar{\Sigma}}\mathbf{\bar{\Sigma} Z_R}=\mathbf{U}\mathbf{V}^H
\end{equation}  
where both $\mathbf{U}$ and $\mathbf{V}$ are also Vandermonde matrices. So, the Hankel matrix $\mathscr{H}(\mathbf{x})$ can be decomposed as the product of two low-rank matrices. According to Lemma 8 in \cite{Srebro2004}, the following relation holds true
\begin{equation} \label{eq:nuclearNorm_factorization}
    \|\mathscr{H}(\mathbf{x})\|_\ast = \min\limits_{\substack{\tilde{\mathbf{U}},\tilde{\mathbf{V}} \\ \mathscr{H}(\mathbf{x})= \mathbf{\tilde{U}\tilde{V}}^H } } \frac{1}{2} \left( \|\tilde{\mathbf{U}}\|_F^2 +\|\tilde{\mathbf{V}}\|_F^2 \right) 
\end{equation}
where $\tilde{\mathbf{U}}$ and $\tilde{\mathbf{V}}$ has no constraint in their sizes; thus, they could be different from the $\mathbf{U}$ and $\mathbf{V}$ in \eqref{eq:Hankel_X_UV}. 

Taking advantage of \eqref{eq:nuclearNorm_factorization} as a factorization of the nuclear norm, the optimization problem \eqref{eq:sigSep_vecOpt_x_i}  can be rewritten as 
\begin{align}\label{eq:sigSep_vecOpt_factor_nuNorm_relax}
\left\{\mathbf{x,i, U, V}\right\} = \arg \min_{\mathbf{U},\mathbf{V},\mathbf{x,i}} \quad & \frac{1}{2}  \left(\left\|\mathbf{U}\right\|_F^2 + \left\|\mathbf{V}\right\|_F^2 \right) + \tau  \left\|\mathbf{i}\right\|_1\\
\text{s.t.} \quad
& \left\|\mathbf{y-x-i}\right\|_2^2  \leq \varepsilon \nonumber\\
&  \mathscr{H}(\mathbf{x}) = \mathbf{UV}^H \nonumber 
\end{align} 
Note that for simplicity of notation  $\tilde{\mathbf{U}}$ and $\tilde{\mathbf{V}}$ are replaced by $\mathbf{U}$ and $\mathbf{V}$ in \eqref{eq:sigSep_vecOpt_factor_nuNorm_relax}.
This problem is much easier to address in terms of the optimization process. It can be solved with the ADMM iterative scheme, which would result in an SVD-free optimization algorithm. 

\section{Sparse and Low-rank Decomposition of the Hankel Matrix} \label{sec:Approach}

\subsection{Algorithm Derivation}
In this section, an ADMM-based optimization approach is derived to address the optimization problem \eqref{eq:sigSep_vecOpt_factor_nuNorm_relax}. Its augmented Lagrangian function can be written as 
\begin{equation}
\begin{aligned}
&\mathcal{L}(\mathbf{U},\mathbf{V},\mathbf{x},\mathbf{i},\mathbf{p},\mathbf{Q}) \nonumber\\
& \quad = \frac{1}{2} (\left\|\mathbf{U}\right\|_F^2 + \left\|\mathbf{V}\right\|_F^2  ) + \tau  \left\|\mathbf{i}\right\|_1  \nonumber\\
&\quad + \frac{\beta}{2}\left\|\mathbf{y-x-i}\right\|_2^2 + \langle \mathbf{p,y-x-i} \rangle  \nonumber\\
&\quad +  \frac{\mu}{2}\left\|\mathscr{H}(\mathbf{x})-\mathbf{UV}^H\right\|_F^2 +\langle \mathbf{Q},\mathscr{H}(\mathbf{x})-\mathbf{UV}^H \rangle.
\end{aligned}
\end{equation}
where $\mathbf{p}$ and $\mathbf{Q}$ are multipliers, and $\beta$ and $\mu$ are regularization parameters.  Combining the linear and quadratic terms in the augmented Lagrangian function, the scaled dual form can be expressed as 
\begin{equation}\label{eq:Largrangian function}
   \begin{aligned} 
    \mathcal{L}(\mathbf{U,V,x,i,p,Q})
   & = \frac{1}{2} (\left\|\mathbf{U}\right\|_F^2 + \left\|\mathbf{V}\right\|_F^2  ) + \tau  \left\|\mathbf{i}\right\|_1\\
    & + \frac{\beta}{2}\left\|\mathbf{y-x-i}+\frac{1}{\beta}\mathbf{p}\right\|_2^2 \\
    & +  \frac{\mu}{2}\left\|\mathscr{H}(\mathbf{x})-\mathbf{UV}^H+\frac{1}{\mu}\mathbf{Q}\right\|_F^2 .
    \end{aligned} 
\end{equation}

Based on the augmented Lagrangian function \eqref{eq:Largrangian function}, the minimization optimization problem in \eqref{eq:sigSep_vecOpt_factor_nuNorm_relax} can be solved using the following ADMM iterative scheme:
\begin{align}
\mathbf{x}^{(k+1)} &=  \arg\min_{\mathbf{x}} \mathcal{L}(\mathbf{U}^{(k)},\mathbf{V}^{(k)},\mathbf{x},\mathbf{i}^{(k)},\mathbf{p}^{(k)},\mathbf{Q}^{(k)}) \label{Eq: Update_x} \\
\mathbf{i}^{(k+1)} &= \arg  \min_i \mathcal{L}(\mathbf{U}^{(k)},\mathbf{V}^{(k)},\mathbf{x}^{(k+1)},\mathbf{i},\mathbf{p}^{(k)},\mathbf{Q}^{(k)}) \label{Eq: Update_i} \\
\mathbf{U}^{(k+1)}  &= \arg \min_{\mathbf{U}} \mathcal{L}(\mathbf{U},\mathbf{V}^{(k)},\mathbf{x}^{(k+1)},\mathbf{i}^{(k+1)},\mathbf{p}^{(k)},\mathbf{Q}^{(k)}) \label{Eq: Update_U} \\
\mathbf{V}^{(k+1)} &= \arg   \min_{\mathbf{V}} \mathcal{L}(\mathbf{U}^{(k+1)},\mathbf{V},\mathbf{x}^{(k+1)},\mathbf{i}^{(k+1)},\mathbf{p}^{(k)},\mathbf{Q}^{(k)}) \label{Eq: Update_V} \\
\mathbf{p}^{(k+1)} &= \mathbf{p}^{(k)} +\beta\left(\mathbf{y}-\mathbf{x}^{(k+1)}-\mathbf{i}^{(k+1)} \right) \label{Eq: Update_p} \\
\mathbf{Q}^{(k+1)} &= \mathbf{Q}^{(k)} +\mu\left[\mathscr{H}\left(\mathbf{x}^{(k+1)}\right)-\mathbf{U}^{(k+1)}\left(\mathbf{V}^{(k+1)}\right)^H\right]  \label{Eq: Update_Q}
\end{align}

In this iterative scheme, the four simple optimization problems \eqref{Eq: Update_x}-\eqref{Eq: Update_V} have to be addressed, which are discussed below in detail. 
\vspace{2mm}
~\\
(1) Update $\mathbf{x}$

To update the variable $\mathbf{x}$, the optimization problem in \eqref{Eq: Update_x} can be written as 
\begin{multline}
\mathbf{x} = \arg \min_\mathbf{x} \ \left[\frac{\beta}{2}\left\|\mathbf{y-x-i}^{(k)}+\frac{1}{\beta} \mathbf{p}^{(k)} \right\|_2^2 \right. \\
 \left. + \frac{\mu}{2}\left\|\mathscr{H}(\mathbf{x})-\mathbf{U}^{(k)} (\mathbf{V}^{(k)})^H+\frac{1}{\mu}\mathbf{Q}^{(k)}\right\|_F^2 \right]
\end{multline}
Taking the first derivative, one can get the updated $\mathbf{x}$ as
\begin{multline} \label{Eq: Update x 2}
    \mathbf{x}^{(k+1)} = \frac{1}{\mu+\beta}\left\{ \beta \left(\mathbf{y-i}^{(k)}+\frac{1}{\beta} \mathbf{p}^{(k)}\right) \right. \\
    \left. + \mu \mathscr{H}^{\dagger} \left[ \mathbf{U}^{(k)} \left( \mathbf{V}^{(k)} \right)^H - \frac{1}{\mu} \mathbf{Q}^{(k)} \right] \right\} 
\end{multline}
where  $\mathscr{H}^{\dagger}(\mathbf{X})=[\mathbf{X}(:,1)^T, \mathbf{X}(M,2:N)]^T$ constructs a column vector $\tilde{\mathbf{x}}$ by picking from a Hankel matrix $\mathbf{X} \in \mathbb{C}^{M\times N} $ all the entries in the first column $\mathbf{X}(:,1)$ and all the entries in the last row except the first one, i.e., $\mathbf{X}(M,2:N)$, which denotes the Penrose-Moore pseudoinverse of  $\mathscr{H}(\tilde{\mathbf{x}})$.

~\\
(2) Update $\mathbf{i}$

The corresponding optimization problem in \eqref{Eq: Update_i} can be explicitly written as
\begin{equation} \label{Eq: Update i 2}
\mathbf{i}^{(k+1)} = \arg \min_\mathbf{i} \ \tau  \left\|\mathbf{i}\right\|_1 + \frac{\beta}{2} \left\|\mathbf{y-x}^{(k+1)}-\mathbf{i}+\frac{1}{\beta} \mathbf{p}^{(k)}\right\|_2^2 ,
\end{equation} 
which is a typical Lasso (Least absolute shrinkage and selection operator) regression problem. Its solution is given by
\begin{equation} \label{eq:Update_i_express_explicit}
\mathbf{i}^{(k+1)} =  \mathcal{S}_{\tau/\beta} \left( \mathbf{y-x}^{(k+1)}+ \frac{1}{\beta} \mathbf{p}^{(k)} \right)
\end{equation}
where $\mathcal{S}_\lambda (x) = e^{j\text{arg}(x)}(|x|-\lambda)_{+} $ denotes the complex soft thresholding operator.\\
~\\
(3) Update $\mathbf{U}$

The optimization problem in \eqref{Eq: Update_U} for $\mathbf{U}$ update can be simply expressed as
\begin{equation} \label{Eq: Update U 2}
\mathbf{U}^{(k+1)}  = \arg \min_{\mathbf{U}} \frac{1}{2} \left\|\mathbf{U}\right\|_F^2 + \frac{\mu}{2} \left\|\mathscr{H}(\mathbf{x}^{(k+1)})-\mathbf{U}(\mathbf{V}^{(k)})^H\right\|_F^2  
\end{equation}
By taking the first derivative, the minimizer can be obtained
\begin{multline} \label{eq:Update_U_express_explicit}
\mathbf{U}^{(k+1)} = \mu \left( \mathscr{H}(\mathbf{x}^{(k+1)}) + \frac{1}{\mu} \mathbf{Q}^{(k)} \right) \mathbf{V}^{(k)} \\
\cdot \left( \mathbf{I} + \mu(\mathbf{V}^{(k)})^H \mathbf{V}^{(k)} \right)^{-1}
\end{multline}

~\\
(4) Update $\mathbf{V}$

By performing the similar manipulations to update $\mathbf{U}$, one can get the solution to \eqref{Eq: Update_V} to update $\mathbf{V}$, which is given by
\begin{align} \label{Eq: Update V 2} 
	\mathbf{V}^{(k+1)} &= \arg \min_{\mathbf{V}} \left( \frac{1}{2}\left\|\mathbf{V}\right\|_F^2 \right.  \nonumber\\
   &\qquad\qquad	\left. + \frac{\mu}{2}\left\|\mathscr{H} \left(\mathbf{x}^{(k+1)} \right)-\mathbf{U}^{(k+1)}\mathbf{V}^H\right\|_F^2 \right) \nonumber  \\
	 &= \mu \left( \mathscr{H} \left( \mathbf{x}^{(k+1)} \right) + \frac{1}{\mu} \mathbf{Q}^{(k)} \right)^H \mathbf{U}^{(k+1)} \nonumber \\
	&\qquad\qquad\qquad \cdot \left( \mathbf{I} + \mu(\mathbf{U}^{(k+1)})^H\mathbf{U}^{(k+1)} \right)^{-1}
\end{align} 	  

Based on \eqref{Eq: Update x 2}, \eqref{eq:Update_i_express_explicit}, \eqref{eq:Update_U_express_explicit}, \eqref{Eq: Update V 2}, \eqref{Eq: Update_p} and \eqref{Eq: Update_Q}, one can successively update $\mathbf{x}, \mathbf{i},\mathbf{U},\mathbf{V}, \mathbf{p}$ and $\mathbf{Q}$ in a loop. After several iterations, $\mathbf{x}$ and $\mathbf{i}$ are recovered from the measurements with an expected relative error $\delta$. 

Note in the iterative updates three regularization parameters $\beta$, $\mu$ and $\tau$ are involved and they should be set before starting the iterations. The choice of their values generally depends on specific problems. In implementation, to accelerate the convergence of the algorithm, $\beta$ and $\mu$ could be gradually increased to improve the regularization penalties of the data consistency and the signal model constraint, respectively. Empirically, the hyper-parameter $\mu$ could be increased by a factor of $k_\mu(\geq 1)$ to smoothly tight the constraint of signal model while $\beta$ grows $k_\beta(\geq 1)$ times every a few (e.g., $L$) iterations to enhance the consistency between the measured data and recovered signal. The overall algorithm is summarized in Algorithm~\ref{alg:Algorithm_HankelDecom}.

\begin{algorithm}[!b]
\SetAlgoNoLine
\KwData{signal $\mathbf{y}$; initialization of $\mathbf{x}$, $\mathbf{i}$, $\mathbf{p}$, $\mathbf{U}$, $\mathbf{V}$, and $\mathbf{Q}$; Set the relative error $\delta$.  \\
\hspace{8.5mm}	Set the count of iterations $k=0$. \\
\hspace{8.5mm} Set the (initial) values of the regularization\\\quad parameters $\mu$, $\beta$ and $\tau$. }   
\KwResult{$\mathbf{x}$ and $\mathbf{i}$}
\vspace{1mm}
\While{$\|\mathbf{y-x-i}\|_2 \geq \delta \|\mathbf{y}\|_2$}{
	$k = k+1$;

	\If{$0\equiv k \pmod L $}{$\beta\leftarrow \beta \cdot k_\beta $  \tcp*{$\beta$ increased every $L$ iterations}}

1) Update $\mathbf{x}$ by using Equ. \eqref{Eq: Update x 2}; \\
\vspace{1mm}

2) Update $\mathbf{i}$ by using Equ. \eqref{Eq: Update i 2}; \\
\vspace{1mm}
3) Update $\mathbf{U}$ by using Equ. \eqref{Eq: Update U 2}; \\
\vspace{1mm}
4) Update $\mathbf{V}$ by using Equ. \eqref{Eq: Update V 2}; \\
\vspace{1mm}

5) Update $\mathbf{p}$ by using Equ.~\eqref{Eq: Update_p}; \\
\vspace{1mm}

6) Update $\mathbf{Q}$ by using Equ.~\eqref{Eq: Update_Q}; \\
\vspace{1mm}

$\mu \leftarrow \mu \cdot k_\mu$  \tcp*{$\mu$ increased every iteration}

} 
\caption{Interference mitigation with sparse and low-rank decomposition of the Hankel matrix}
\label{alg:Algorithm_HankelDecom}
\end{algorithm}

\section{Numerical Simulations} \label{sec:Simulation_results}

{ \renewcommand{\arraystretch}{1.05}
	\begin{table}[!t]
		\centering
		\caption{System parameters for FMCW radar simulations}
		\label{tab:Simu_parameters}
		\begin{tabular}{ll}
			\toprule
			\textbf{Parameter}         & \textbf{Value} \\ \midrule
			Center frequency           & $3\,\mathrm{GHz}$          \\ 
			Sweep time of FMCW signal  & $400\,\mathrm{\mu s}$         \\ 
			Signal bandwidth           & $40\,\mathrm{MHz}$         \\ 
			Cutoff frequency of the LPF & $5.33\,\mathrm{MHz}$           \\ 
			Sampling frequency         & $12\,\mathrm{MHz}$         \\ 
			Waveform                   & Up-sweep chirp \\ \bottomrule
		\end{tabular}
	\end{table}
}

 In this section, numerical simulations are presented to demonstrate the interference mitigation performance of the proposed approach. Meanwhile, the results are also compared with that of three state-of-the-art approaches: the adaptive noise canceller (ANC) \cite{Jin2019ANC}, SALSA-based approach \cite{Uysal2019} and RPCA-based method \cite{Candes2011}, where the first one is a filtering approach while the latter two mitigate interferences through signal separation. Note that for the SALSA-based approach and the RPCA-based method, the formulations in \eqref{eq:optimProblem_SALSAR} and \eqref{eq:sigSep_matOpt_nuNorm_L1} are respectively used for interference mitigation below.   

\subsection{Evaluation metric}
To quantitatively evaluate the interference mitigation performance of the proposed approach and facilitate its comparison with other methods in the following, we first introduce two metrics: signal-to-interference-plus-noise ratio (SINR) and correlation coefficient $\rho$. The SINR is defined as
\begin{equation} \label{eq:SINR}
\text{SINR} = 20\log_{10} \frac{\|\mathbf{s}\|_2}{\|\mathbf{i} + \mathbf{n} \|_2}
\end{equation} 
where $\mathbf{s}$ is the interference- and noise-free (or reference) signal, and $\mathbf{i}$ and $\mathbf{n}$ are the interference and noise, respectively. For the recovered signal $\hat{\mathbf{s}}$ after interference mitigation, the SINR is obtained by replacing the term $\mathbf{i} + \mathbf{n}$ in \eqref{eq:SINR} with $\mathbf{s} - \hat{\mathbf{s}}$, which is the reciprocal of the relative error of the recovered signal in decibel scale. So the higher the SINR of the recovered signal, the better the interference mitigation. To avoid possible confusion, in the following we use $\text{SINR}_0$ and SINR to denote the signal to noise ratios of a signal before and after IM processing, respectively. The correlation coefficient between the signal $\mathbf{s}$ and its recovered counterpart $\hat{\mathbf{s}}$ is defined as 
\begin{equation}
\rho = \frac{\hat{\mathbf{s}}^H \mathbf{s}}{\left \|\mathbf{s}\right\|_2 \cdot \left \|\hat{\mathbf{s}}\right\|_2}.
\end{equation} 
Generally, $\rho$ is a complex number with its modulus $0 \leq |\rho| \leq 1$. A larger modulus of $\rho$ indicates a higher correlation between $\hat{\mathbf{s}}$ and $\mathbf{s}$ while its phase represents a relative phase different between them.  

\subsection{Point target Simulations} \label{sec:point_target_simu}

\begin{figure*}[!t]
	\centering
	\vspace{-3mm}
	\subfloat[]{
		\includegraphics[width=0.4\textwidth]{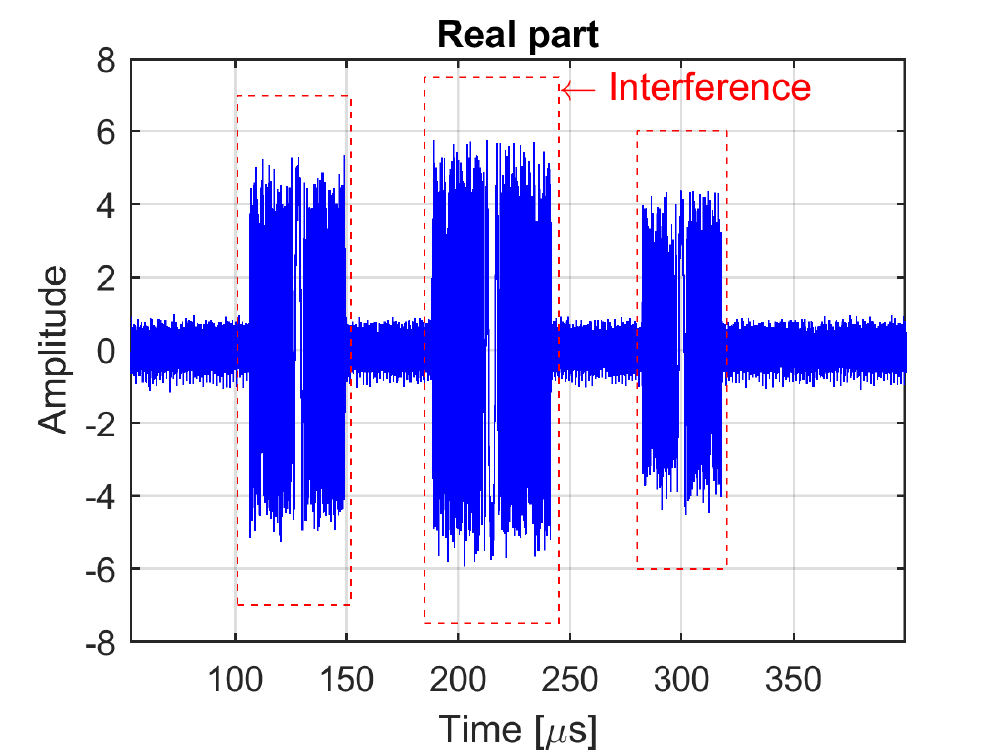}
		\label{fig:simu_sig_full}
	}
	\subfloat[]{
		\includegraphics[width=0.45\textwidth]{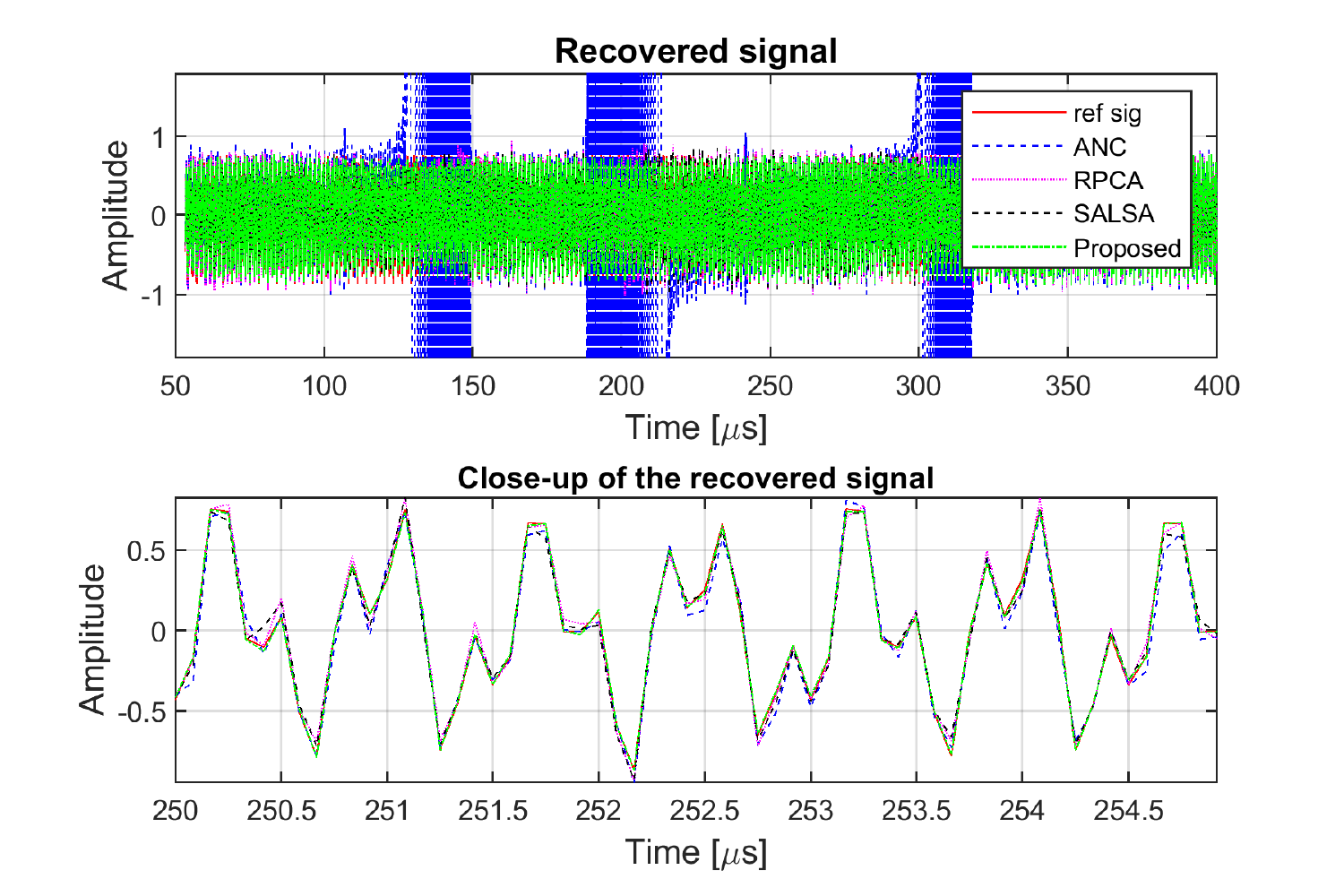}
		\label{fig:simu_extrac_usable_sig_SNR_15}
	}
	\caption{Illustration of interference mitigation for FMCW signals contaminated by multiple interferences. \protect\subref{fig:simu_sig_full} the real part of the interference-contaminated beat signal, \protect\subref{fig:simu_extrac_usable_sig_SNR_15} the recovered signals after interference mitigation with the four methods, where the upper panel shows the recovered beat signal in a whole sweep and the lower panel gives a close-up of the signal segment within the interval $[250,255]\mu s$. }
	\label{fig:simu_IntMitig_SNR_15}
\end{figure*}

{
	\renewcommand{\arraystretch}{1.05}
\begin{table}[!t]
	\vspace{-3mm}
	\centering
	\caption{SINR and correlation coefficient of the recovered signals for point target simulation}
	\label{tab:point_quantitative_metrics}
	\begin{tabular}{lccc}
		\toprule
		& \textbf{SINR[dB]} & $\mathbf{\rho}$ & \begin{tabular}{c}
			\textbf{running} \\ \textbf{time[s]}
		\end{tabular} \\ \midrule
		ANC      &  $-9.69$ &  $0.3394e^{-j0.0471}$  &  0.0013 \\
		SALSA    &  $11.70$ &  $0.9669e^{-j0.0164}$  & 15.2008  \\
		RPCA     &  $14.20$ &  $0.9815e^{-j0.0085}$  & 1050.0966   \\
		Proposed &  $29.95$ &  $0.9995e^{-j0.0006}$  & 70.6254   \\ \bottomrule
	\end{tabular}
\end{table}
}

\begin{figure*}
	\centering
	\vspace{-3mm}
	\subfloat[]{ \hspace{-3mm}
		\includegraphics[width=0.4\textwidth]{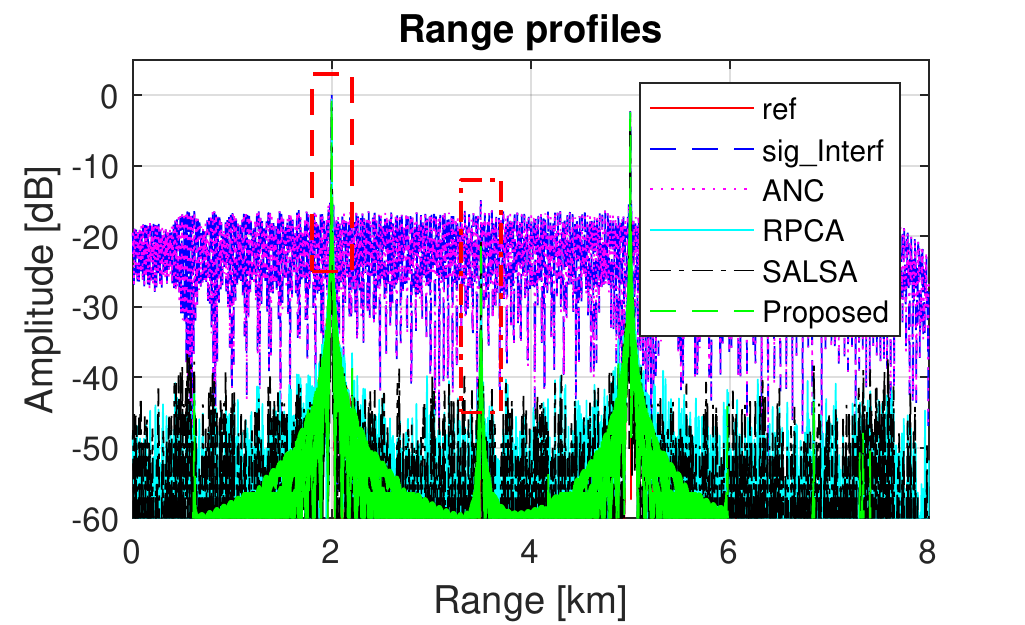}	
		\label{fig:simu_PtTar_RP_SNR_15}
	}
	\subfloat[]{ \hspace{-7mm}
		\includegraphics[width=0.32\textwidth]{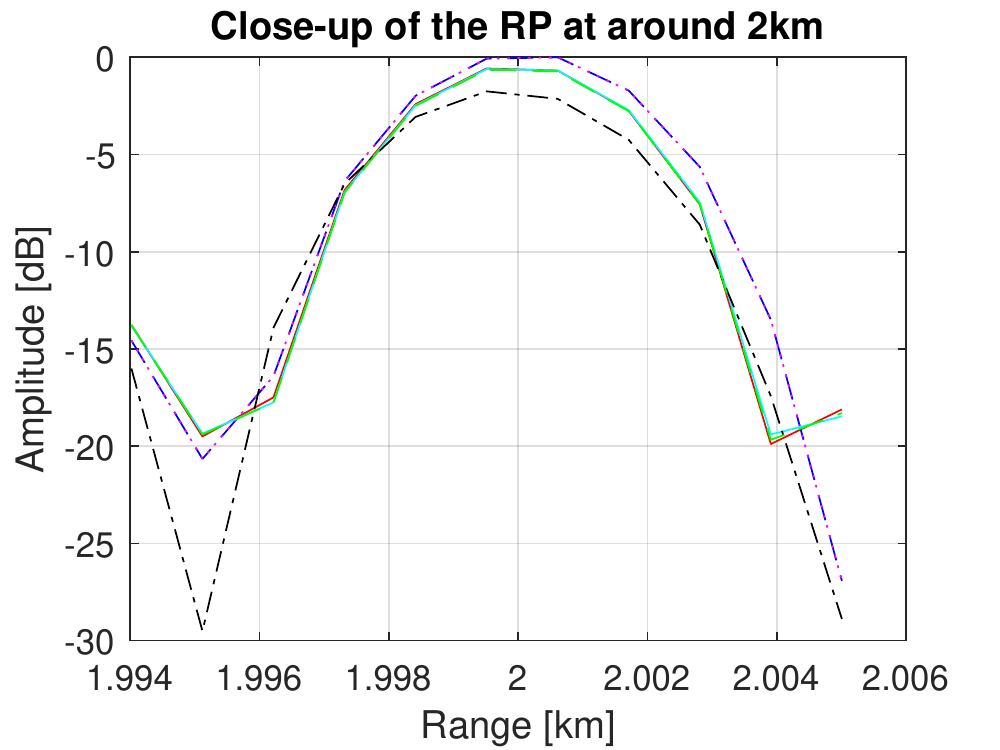}
		\label{fig:simu_PtTar_RP_SNR_15_Rd_2}
	}
	\subfloat[]{ \hspace{-6mm}
		\includegraphics[width=0.32\textwidth]{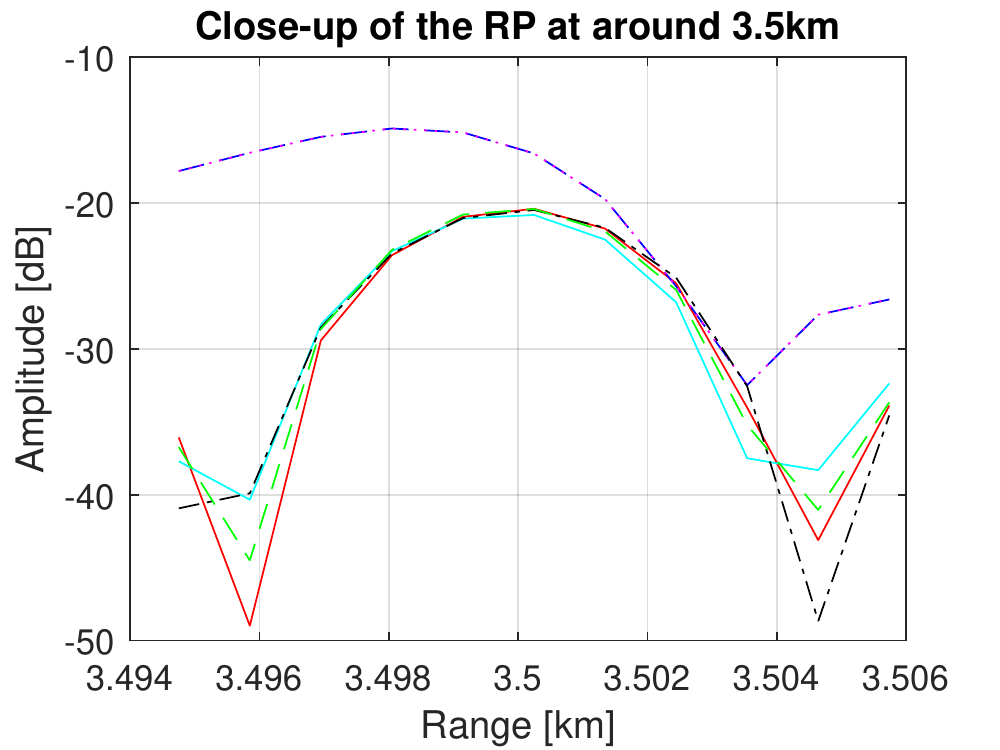}
		\label{fig:simu_PtTar_RP_SNR_15_Rd_35}
	}
	\caption{Range profile of the point targets. \protect\subref{fig:simu_PtTar_RP_SNR_15} shows the range profiles of targets obtained before and after interference mitigation with the four methods. \protect\subref{fig:simu_PtTar_RP_SNR_15_Rd_2} and \protect\subref{fig:simu_PtTar_RP_SNR_15_Rd_35} are the zoomed-in views of the parts indicated by the rectangles in \protect\subref{fig:simu_PtTar_RP_SNR_15}, i.e., the range profiles of targets at the distance of $2\,\mathrm{km}$ and $3.5\,\mathrm{km}$, respectively.}
	\label{fig:simu_PtTar_RP}
\end{figure*}

{ \renewcommand{\arraystretch}{1.05}
	\begin{table*}[!t]
		\centering
		\caption{Quantitative comparison of the SINR and correlation coefficient of the recovered signals by the ANC, SALSA-, RPCA-based methods and the Proposed interference mitigation approach.}
		\label{tab:simu_quantitative_compare}
		\begin{tabular}{@{}lccccccccc@{}}
			\toprule
			\multirow{2}{*}{} & \multicolumn{3}{c}{\boldmath$\textbf{SINR}_0=-20\,\mathrm{dB}$} & \multicolumn{3}{c}{\boldmath$\textbf{SINR}_0=-10\,\mathrm{dB}$} & \multicolumn{3}{c}{\boldmath $\textbf{SINR}_0=0\,\mathrm{dB}$} \\ \cmidrule{2-10} 
			& SINR[dB]         & $\rho$       & \begin{tabular}{@{}c@{}}
				running \\ time[s]
			\end{tabular}   &  SINR[dB]          & $\rho$      & \begin{tabular}{@{}c@{}}
			running \\ time[s]
		\end{tabular} & SINR[dB]         & $\rho$ & \begin{tabular}{@{}c@{}}
		running \\ time[s]
	    \end{tabular}     \\ \midrule
			ANC               & $-16.9939$          & $0.1474e^{-j0.0098}$  & $\mathbf{0.0052}$       & $-7.0345$       & $0.4089e^{j0.0149}$   & $\mathbf{0.0012}$    & $3.0288$       & $0.8161e^{-j0.0177}$   & $\mathbf{0.0010}$   \\ 
			SALSA             & $-7.4778$        & $0.3484e^{-j0.0081}$   & $15.4845$       & $12.4891$       & $0.9752e^{j0.0064}$    & $15.1590$    & $12.7830$      & $0.9774e^{-j0.0057}$   & $15.5027$   \\ 
			RPCA              & $-11.1290$        & $0.2545e^{-j0.0654}$    & $862.2684$     & $15.2970$        & $0.9852e^{j0.0091}$   & $836.8579$     & $16.6967$      & $0.9893e^{-j0.0066}$   & $318.6887$   \\ 
			Proposed          & $\mathbf{24.2185}$         & $\mathbf{0.9987e^{j0.0042}}$   & $50.5626$      & $\mathbf{27.7170}$       & $\mathbf{0.9992e^{j0.0040}}$   & $78.7157$    & $\mathbf{28.4259}$      &  $\mathbf{0.9994e^{-j0.0022}}$   & $70.0923$    \\ \bottomrule
		\end{tabular}
	\end{table*}
}

Firstly, an FMCW radar contaminated by multiple FMCW interferences is considered. The parameters of the FMCW radar system used for the simulations are listed in Table~\ref{tab:Simu_parameters}. Three point targets are placed at a distance of $2\,\mathrm{km}$, $3.5\,\mathrm{km}$ and $5\,\mathrm{km}$, respectively, away from the radar system. 

According to Table~\ref{tab:Simu_parameters}, the radar transmits FMCW waveform with the sweep slope of $K_r=1\times10^5\,\mathrm{MHz/s}$. Assume that it is interfered by three FMCW aggressor radars with sweep slopes of $3K_r$, $-2K_r$ and $-1.5K_r$. The received signals are mixed with the transmitted FMCW waveform for dechirping and then pass through an anti-aliasing low-pass filter (LPF) with the cut-off frequency of 5.33\,MHz. The acquired beat signal of the point targets is shown in  Fig.~\ref{fig:simu_IntMitig_SNR_15}\subref{fig:simu_sig_full}, where about 38\% of the signal samples are contaminated by the three strong FMCW interferences. Meanwhile, white Gaussian noise is added with the signal to noise ratio (SNR) of $15\,\mathrm{dB}$ to consider the system thermal noise and the measurement errors. As a result, the synthesized signal has a $\text{SINR}_0$ of $-12.7\,\mathrm{dB}$. 

Taking advantage of the proposed interference mitigation approach as well as three other state-of-the-art methods, i.e., ANC, SALSA- and RPCA-based methods, the useful beat signals are recovered, which are shown in Fig.~\ref{fig:simu_IntMitig_SNR_15}\subref{fig:simu_extrac_usable_sig_SNR_15}. For comparison, the reference signal is also presented. In the implementation, a filter of length 50 was used for the ANC method. For the morphological component analysis formulation in \eqref{eq:optimProblem_SALSAR}, the regularization parameters $\lambda=0.85$ and $\mu=1$ (accounting for the constraint of data consistency) \cite{Uysal2019} were used for the SALSA-based method where the STFT utilized a 128-sample window with 122 samples of overlap between neighboring segments and a 128-point FFT. The maximum number of iteration for the SALSA method was set to be 1000. For the RPCA-based method, the regularization parameter $\tau$ was set as $1/\sqrt{d_\mathbf{X}}$ where $d_\mathbf{X}$ is the maximum of the lengths of the row and column dimensions of the matrix $\mathbf{X}$ and $\mu=0.05$ was chosen for the data consistency constraint. Regarding the proposed approach, $\tau=0.02$ was used, and $\beta$ and $\mu=0.02$ were initialized with 0.1 and 0.02, respectively; then, $\beta$ was increased by a factor of $k_\beta=1.6$ every $L=10$ iterations and $\mu$ grew $k_\mu=1.2$ times in each iteration. Moreover, the termination of both the RPCA-based method and the proposed approach were determined by setting $\delta=1\times10^{-6}$. Note that the parameters related to each method have been empirically tuned for their optimal performance (without explicit statement, these (hyper-)parameters will be used for all the simulations below). 

From Fig.~\ref{fig:simu_IntMitig_SNR_15}\subref{fig:simu_extrac_usable_sig_SNR_15}, all the SALSA-, PRCA-based methods and the proposed approach successfully mitigate the interferences and outperform the ANC. By constrast, ANC suppresses only half of the interferences compared to the signal in Fig.~\ref{fig:simu_IntMitig_SNR_15}\subref{fig:simu_sig_full}. As the ANC assumes that the spectra of interferences in the positive and negative frequency bands are conjugate symmetric while the useful beat signals are only located in the positive spectrum, which generally is not fulfilled in practice, the negative spectrum is filtered out and then utilized as a reference to eliminate the interference in the positive spectra. As the conjugate symmetry of the spectrum of interferences is not satisfied in this case, the gain of the interference mitigation of ANC is mainly attributed to the spectrum filtering; thus, one can see that half of the interferences are suppressed by ANC compared to Fig.~\ref{fig:simu_IntMitig_SNR_15}\subref{fig:simu_sig_full}. From the bottom panel in Fig.~\ref{fig:simu_IntMitig_SNR_15}\subref{fig:simu_extrac_usable_sig_SNR_15}, visually the recovered beat signal obtained with the proposed approach shows the best agreement with the reference one. For quantitative evaluation, the SINRs and $\rho$'s of the recovered signals with the ANC, SALSA, RPCA and the proposed approaches are shown in Table~\ref{tab:point_quantitative_metrics}. Compared to the three existing methods, the proposed approach significantly supresses the interference, improving the SINR and $\rho$ of the recovered signal relative to the reference.      

Applying the FFT operation of the recovered signal with respect to time, the range profile of targets is obtained, as shown in Fig.~\ref{fig:simu_PtTar_RP}\subref{fig:simu_PtTar_RP_SNR_15}. For comparison, the range profiles of the reference signal (i.e., denoted by ``ref'') and the beat signal before interference mitigation (i.e., ``sig\_Interf'') are illustrated as well. One can see that the ANC only partially removes the strong interference and the remaining interference still causes very high ``noise'' floor in the resultant range profile. By contrast, the SALSA-, RPCA-based methods and the proposed approach all substantially mitigate the interferences; thus, they make clearly visible the weak target at $3.5\,\mathrm{km}$ that is almost completely immersed in the high ``noise'' floor caused by strong interferences (see the ``sig\_Interf'' in Fig.~\ref{fig:simu_PtTar_RP}\subref{fig:simu_PtTar_RP_SNR_15}). Moreover, compared to SALSA- and the RPCA-based methods, the proposed approach leads to lower ``noise'' floor in the obtained range profile after interference mitigation, which confirms the high SINR of its recovered beat signal in Fig.~\ref{fig:simu_IntMitig_SNR_15}\subref{fig:simu_extrac_usable_sig_SNR_15}. Meanwhile, the range profiles of targets obtained with the proposed approach show the best agreement with the reference one (see the zoomed-in views in Fig.~\ref{fig:simu_PtTar_RP}\subref{fig:simu_PtTar_RP_SNR_15_Rd_2} and \subref{fig:simu_PtTar_RP_SNR_15_Rd_35}).  

\subsection{Effect of interference strength and duration on the IM performance}

To examine the effects of the strength and the duration of the interference (equivalently, the percentage of the interference-contaminated signal samples) on the interference mitigation performance of the proposed approach, two sets of simulations are performed. Firstly, we keep the duration of the interferences the same as in Fig.~\ref{fig:simu_IntMitig_SNR_15}\subref{fig:simu_sig_full} relative to the duration of the beat signal but change the strengths of the interferences. The synthetic interference-contaminated beat signals with the $\text{SINR}_0$ of $-20\,\mathrm{dB}$, $-10\,\mathrm{dB}$ and $0\,\mathrm{dB}$ were synthesized and
processed with the ANC, SALSA-, RPCA-based methods and the proposed approach. For each $\text{SINR}_0$ case, 20 Monte Carlo simulation runs were performed. The averages of the SINRs and correlation coefficients of the recovered signals are listed in Table~\ref{tab:simu_quantitative_compare}. Compared to the ANC, the SALSA-, RPCA- and the proposed approaches achieve much better interference mitigation performance in terms of both SINR and correlation coefficient in all three $\text{SINR}_0$ cases. Meanwhile, they all noticeably eliminate the interferences and improve the $\text{SINR}$ of the recovered signal and its correlation coefficient relative to the reference signal in the case of  $\text{SINR}_0=-10\,\mathrm{dB}$. Moreover, at all the three $\text{SINR}_0$ levels the  proposed approach constantly outperforms the other three methods, especially, in terms of \text{SINR} of the recovered signal by a significant margin.     

Then, the effects of interference duration on the SINR and $\rho$ are investigated by using interference-contaminated beat signals with various interference durations but a constant $\text{SINR}_0$. The interference-contaminated beat signals were synthesized by simultaneously changing the sweep slopes of the interferences relative to the transmitted FMCW waveform, which leads to various interference durations after dechirping and low-pass filtering operations, and their magnitudes. The $\text{SINR}_0$ of the synthetic beat signals was kept to be almost constant, i.e., about $-16.5\,\mathrm{dB}$. Then, they were processed with the four interference mitigation approaches. For each interference duration case 20 Monte Carlo simulations were run, and the averages of SINRs and correlation coefficients of the recovered signals were computed.

\begin{figure}
    \centering
    \vspace{-5mm}
    \subfloat[]{
    \includegraphics[width=0.4\textwidth]{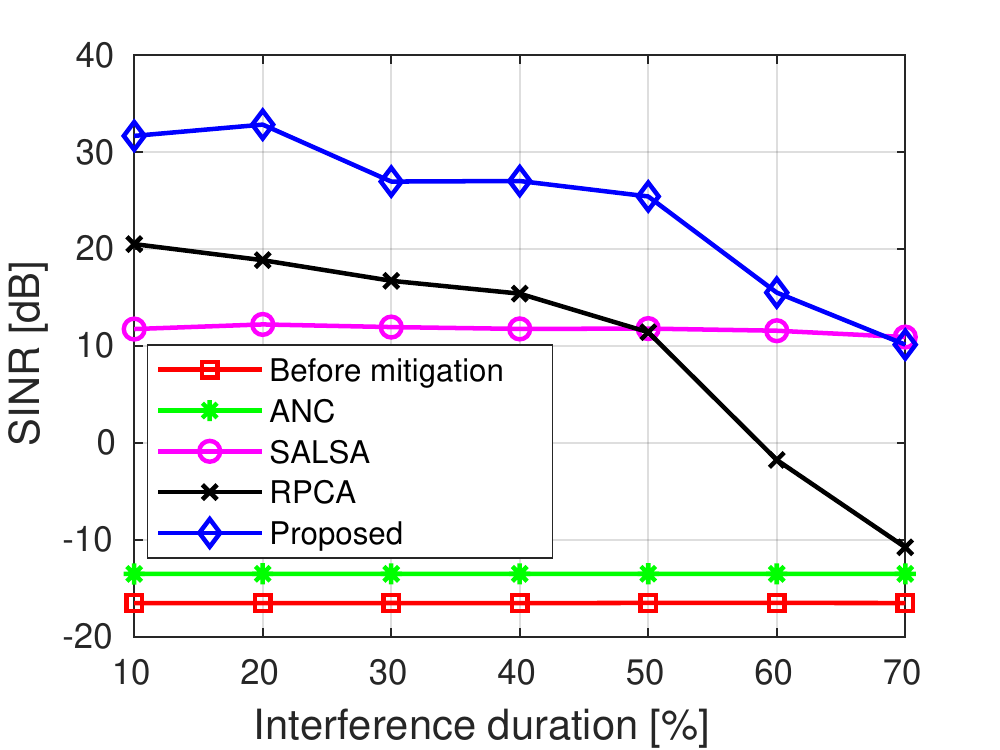}
    \label{fig:simu_interf_mitig_SINT}
    }

	\vspace{-4mm}
    \subfloat[]{
    \includegraphics[width=0.44\textwidth]{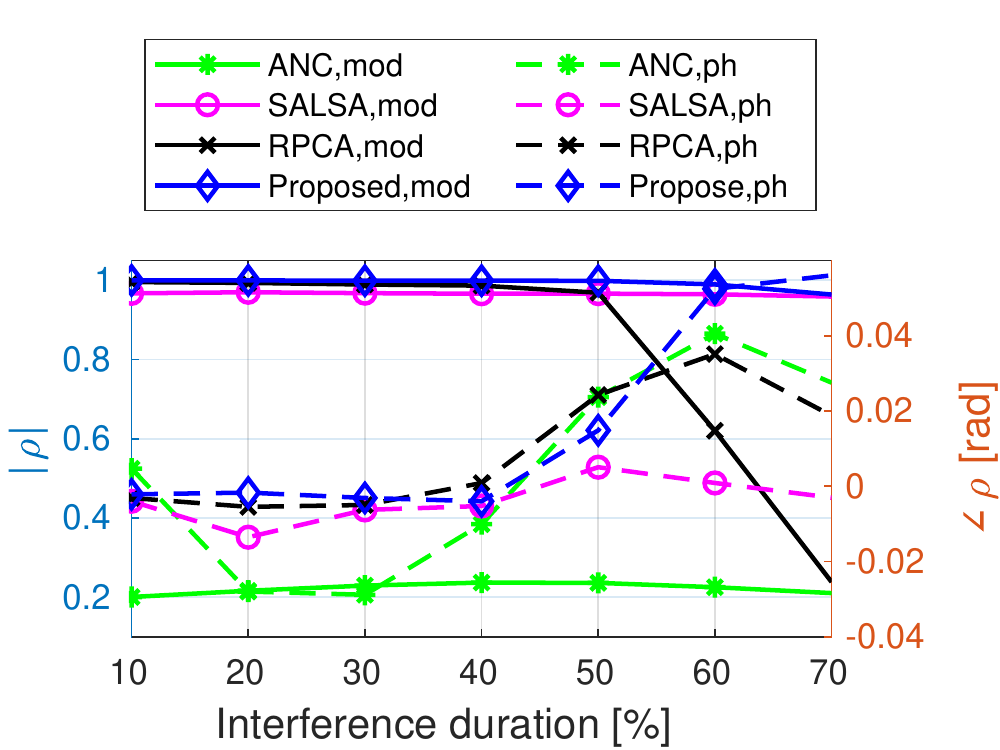}
    \label{fig:simu_interf_mitig_CorrCoef}
    }
    \caption{Impact of the interference duration on the interference mitigation performance. \protect\subref{fig:simu_interf_mitig_SINT} shows the variation of SINR with the interference durations. \protect\subref{fig:simu_interf_mitig_CorrCoef}shows the variation of the correlation coefficients with different interference durations.}
    \label{fig:simu_IntMitig_PerfEvaluate}
\end{figure}

Fig.~\ref{fig:simu_IntMitig_PerfEvaluate} shows the effect of interference duration on the $\text{SINR}$ and $\rho$ of their recovered signals.  As one can see, the ANC and SALSA-based method achieve almost constant $\text{SINR}$ while the performance of the RPCA-based and the proposed approaches degrade with the increase of the interference duration(see Fig.~\ref{fig:simu_IntMitig_PerfEvaluate}\subref{fig:simu_interf_mitig_SINT}). This is because the SALSA-based method exploits the sparse features of useful signal and interferences in the frequency domain and time-frequency domain, respectively. So the time durations of a few interferences have marginal impact on its interference mitigation performance. By contrast, the RPCA-based method and the proposed approach assume that  the interferences are sparse in time. Consequently, a rapid decline of their performance is observed when the interference duration is longer than $50\%$ of the signal duration. Nevertheless, when the interference duration is less than $50\%$, both the RPCA-based method and the proposed approach perform better than the other two methods. Meanwhile, the proposed approach gains about $10\,\mathrm{dB}$ improvement of the SINR of the recovered signal compared to the RPCA-based method. From Fig.~\ref{fig:simu_IntMitig_PerfEvaluate}\subref{fig:simu_interf_mitig_CorrCoef}, one can see that the moduli of $|\rho|$ of the recovered signals by the SALSA-, RPCA-based methods and the proposed approach are all close to 1 and the corresponding phases are near 0 when the interference duration is less than $50\%$; thus, they get more accurate recovered signals than the ANC. Compared to the other three methods, the proposed approach recovers the useful beat signals with the best correlation coefficient $|\rho|$ (i.e., highest moduli and near-zero phase) with respect to the reference ground-truth signal when the interference is sparse in time. However, when the interference duration increases to $50\%$ or more, the assumption of sparse interference in the time domain is violated and then the performance of the proposed approach gradually drops, which results in the decrease of the modulus of $|\rho|$ and the rise of the deviation of its phases from zero.          

\begin{figure}[!t]
    \centering
    \vspace{-4mm}
    \includegraphics[width=0.35\textwidth]{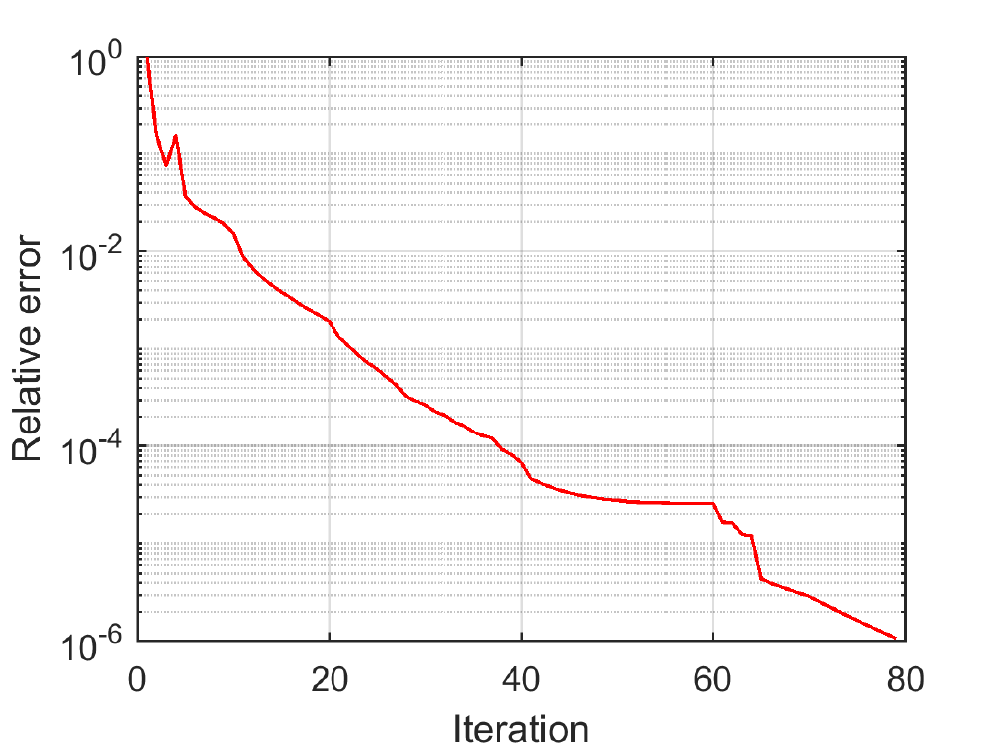}
    \caption{Illustration of the convergence of the proposed IM method with the increase of the running iterations.}
    \label{fig:simu_convergence}
\end{figure}

\begin{figure*}[!t]
\centering
\subfloat{
\includegraphics[width=0.5\textwidth]{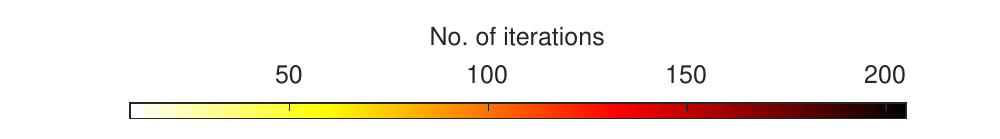}
}

\setcounter{subfigure}{0}
\vspace{-5mm}
\subfloat[]{\hspace{-2mm}
\includegraphics[width = 0.26\textwidth]{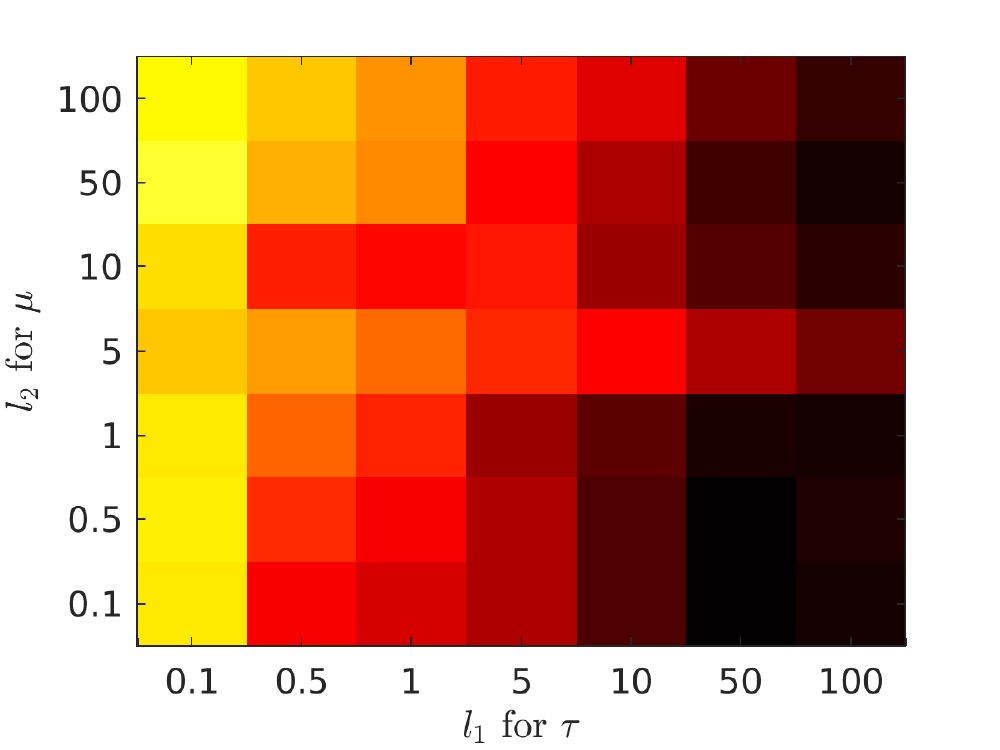}
\label{fig:simu_ParaSel_01}
}
\subfloat[]{\hspace{-4mm}
\includegraphics[width=0.26\textwidth]{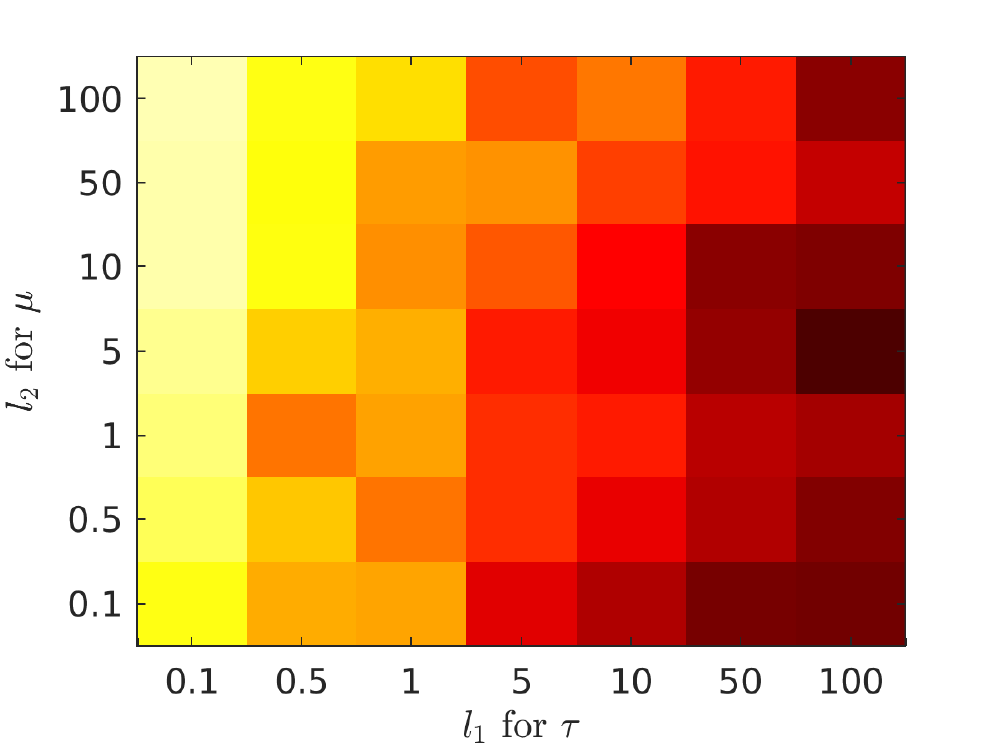}
\label{fig:simu_ParaSel_1}
}
\subfloat[]{\hspace{-4mm}
\includegraphics[width=0.26\textwidth]{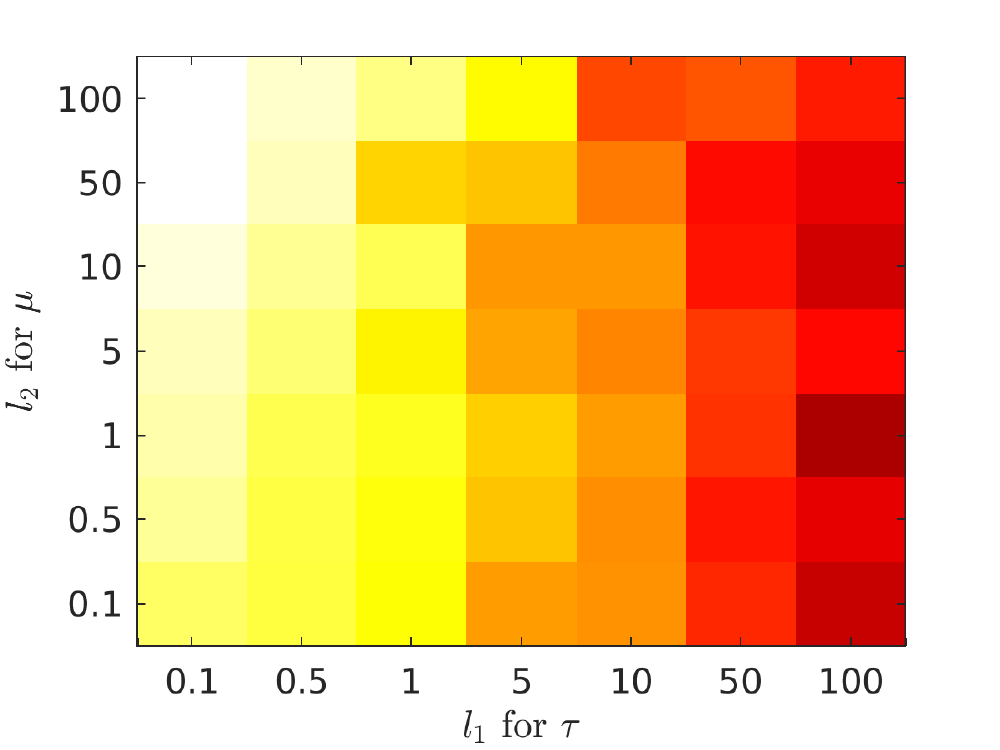}
\label{fig:simu_ParaSel_10}
}
\subfloat[]{\hspace{-4mm}
\includegraphics[width=0.26\textwidth]{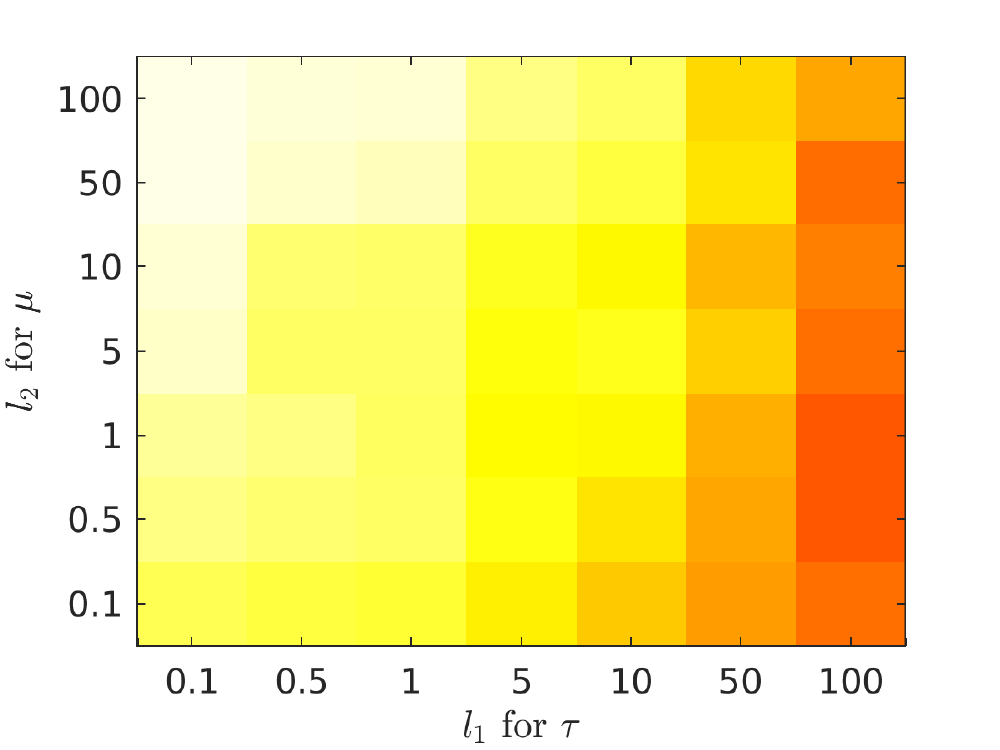}
\label{fig:simu_ParaSel_100}
}
\caption{The number of iterations used by the proposed approach to get the estimated signal with predefined relative error $10^{-6}$ in different combinations of the values of the hyperparameters $\beta$, $\mu$ and $\tau$. $\tau =  \frac{l_1}{\sqrt{\max(m,n)}} $ and $\mu = l_2 \frac{100}{\|\mathbf{Y}\|_2}$ with $l_1,l_2 \in \{0.1, 0.5,1,5,10,50,100\}$. Moreover, in \protect\subref{fig:simu_ParaSel_01} $\beta = \frac{0.1}{10^{\text{SNR}/10}}$, \protect\subref{fig:simu_ParaSel_1} $\beta=\frac{1}{10^{\text{SNR}/10}}$, \protect\subref{fig:simu_ParaSel_10} $\beta = \frac{10}{10^{\text{SNR}/10}}$ and $\beta = \frac{100}{10^{\text{SNR}/10}}$. }
\label{fig:simu_HyperParaSelection}
\end{figure*}

\subsection{Algorithm convergence and hyperparameter selections}
The proposed algorithm is developed based on the ADMM framework. The theoretical analysis of the convergence of ADMM has been discussed in many papers \cite{Boyd2011ADMM,Lin2010}. Considering the limited scope and for the sake of conciseness, we omit the theoretical analysis of the convergence of the developed ADMM-based algortihm but instead illustrate its convergence based on the numerical simulations. Extensive numerical simulations show that the developed algorithm generally converges to a relatively good result in a few tens of iterations by properly choosing the values of hyper-parameters $\beta$, $\mu$ and $\tau$. For the simulation in section~\ref{sec:point_target_simu}, the relative error of the estimation signal with respect to the number of operating iteration of the proposed algorithm is shown in Fig.~\ref{fig:simu_convergence}. Although the relative error had a  fluctuation from iteration 3 to iteration 4, it does not affect the final convergence of the algorithm. After 40 iterations, the relative error has already dropped below $10^{-4}$.    

The hyperparameters $\beta$, $\tau$ and $\mu$ impose the trade-off between data consistancy, time domain sparsity of interferences and spectral sparsity of the useful signals, respectively. For the data consistency constraint, it is natural to use a larger penalty $\beta$ when the SNR of the data is high while a smaller one when the SNR is low. Due to this observation, we recommend to choose $\tau = \frac{l_0}{10^{\text{SNR}/10}}$, where $l_0$ is a constant factor. For the time-domain sparse component, which corresponds to the sparse component in the canonical RPCA problem, the related hyperparameter $\tau$ can take the similar choice suggested in \cite{Candes2011}, i.e., $\tau = \frac{l_1}{\sqrt{\max(m,n)}}$ with $m$ and $n$ are the row and column dimensions of the constructed Hankel matrix, and $l_1$ is a constant factor. Meanwhile, the spectral sparsity constraint of useful signals is characterized by the Frobenius norm of the difference between its Hankel matrix and the product of two low-rank matrices. Thus, similar in \cite{Huang2019},  we recommend $\mu=\frac{100\cdot l_2}{\|\mathbf{Y}\|_2}$, where $\|\mathbf{Y}\|_2$ is the $\ell_2$ norm of the matrix $\mathbf{Y}$, i.e., its largest singular value, and $l_2$ is a constant multiplier. In practice, the multiplication factors $l_0$, $l_1$ and $l_2$ could be tuned for specific signals to get satisfactory results. 
To investigate the effect of the hyperparameters on the convergence of the proposed method, we used the synthetic data in section~\ref{sec:point_target_simu} as an example and took $l_1 \in \{0.1, 1, 10, 100\}$ and $l_1, l_2 \in \{0.1, 0.5, 1, 5, 10, 50, 100\}$ to form various combinations of the three hyperparameters. By setting $k_\mu = 1.2$, $k_\beta = 1.6$, $L=10$ and the relative error $\delta=1\times10^{-6}$, the number of iterations needed to get a desired result with different combinations of the hyperparameters are shown in Fig.~\ref{fig:simu_HyperParaSelection}. It is clear that a larger multiplication factor for $\mu$ and $\beta$ and a smaller one for $\tau$ would make the proposed method converge fast when an estimation of the useful signal with a certain relative error is expected.

\subsection{Computational complexity and time}

Generally, the ANC takes advantage of an adaptive filtering technique for interference mitigation and is a very efficient method. By contrast, the SALSA- and RPCA-based methods as well as the proposed approach are iterative algorithms and their computational complexities depend on the number of iterations in practice. Specifically, the SALSA-based method could exploit the FFT for efficient implementation while the RPCA-based approach has to deal with the SVD of a large matrix in each iteration, which generally is very computationally heavy. By contrast, the proposed approach is an SVD-free algorithm and its computational complexity should be significantly reduced compared to the RPCA-based method. The computational complexity of the proposed approach is mainly determined by the six update steps and the number of iterations. Assuming the maximum rank of $\mathbf{U}$ and $\mathbf{V}$ is $p$, the computational complexity for updating $\mathbf{x}$, $\mathbf{i}$, $\mathbf{U}$, $\mathbf{V}$, $\mathbf{p}$ and $\mathbf{Q}$ in one iteration are $O(N+mn(p+1))$, $O(N)$, $O(mn(1+p)+(m+n+1)p^2+p^3 )$, $O(mn(1+p)+(m+n+1)p^2+p^3 )$, $O(N)$, and $O(mn(p+2))$, respectively. So, in one iteration the computational complexity of the proposed method  is $O(mn(4p+5)+2(m+n+1)p^2 +2p^3)$, which is much smaller than that of the SVD-based method, i.e., $O(mn^2)$ ($m\geq n$) considering the fact $p\ll m,n$.

As an example, the average computation time of the four methods for the simulations in section~\ref{sec:point_target_simu} are listed in Table~\ref{tab:point_quantitative_metrics} and Table~\ref{tab:simu_quantitative_compare}, respectively. All the methods were implemented with MATLAB and run on a PC with an Intel i5-3470 Cnetral Processing Unit (CPU) @ 3.2\,GHz and 8\,GB Random Access Memory (RAM).  It is clear that the ANC is the most efficient method among them and the proposed method is much faster than the RPCA-based approach but slower than the SALSA-based one.

\section{Experimental results}  \label{sec:Experiments}

In this section, the application of the proposed method to real radar data is presented. The same measurement setups and data sets (i.e., the chimney and rain data) as in \cite{JWang2020} are used to verify the proposed approach here. The data was collected with the TU Delft full-polarimetric PARSAX radar by simultaneously transmitting up- and down-chirp signals at the Horizontal (H-) and Vertical (V-) polarization channels for full scattering matrix measurements. The PARSAX radar was designed with a good isolation between receiver and transmitter, i.e., -100\,dB for HH (Horizontally polarized transmission and Horizontally polarized reception) and -85\,dB for VV (Vertically polarized transmission and Vertically polarized reception), and polarization channel isolation better than 30\,dB \cite{Oleg2010PARSAX}.   For convenience, the parameters for experimental measurements are listed in Table \ref{Table: Experimental setup}. The interference of strong co-polarized, i.e., VV-pol signal on the cross-polarized, i.ge., HV-pol (Horizontally polarized transmission and Vertically polarized reception) signals is considered below. 

{ \renewcommand{\arraystretch}{1.05}
	\begin{table}[!t]
		\centering
		\caption{Experimental setup Parameters for Experiment 1 and Experiment 2}
		\label{Table: Experimental setup}
		\begin{tabular}{@{}ll@{}}
			\toprule                     
			\textbf{Parameter}          & \textbf{Value}   \\ \midrule
			Center frequency                    & $3.1315\,\mathrm{GHz}$                            \\ 
			Bandwidth                           & $40\,\mathrm{MHz}$                            \\ 
			Sweep time                          & $1\,\mathrm{ms}$                             \\ 
			Number of samples per sweep                 & 16384                     \\ 
			Maximum range                       & $18.75\,\mathrm{km}$                              \\ 
			Number of sweeps per CPI  & 512 \\ 
			Waveforms
			                            & \begin{tabular}[c]{@{}l@{}l@{}}Up-chirp for H-pol channel;\\ Down-chirp for V-pol channel\end{tabular}  \\ \bottomrule
		\end{tabular}
	\end{table}
}

\begin{figure}[!t]
	\centering
	\subfloat[]{
		\includegraphics[width=0.48\textwidth]{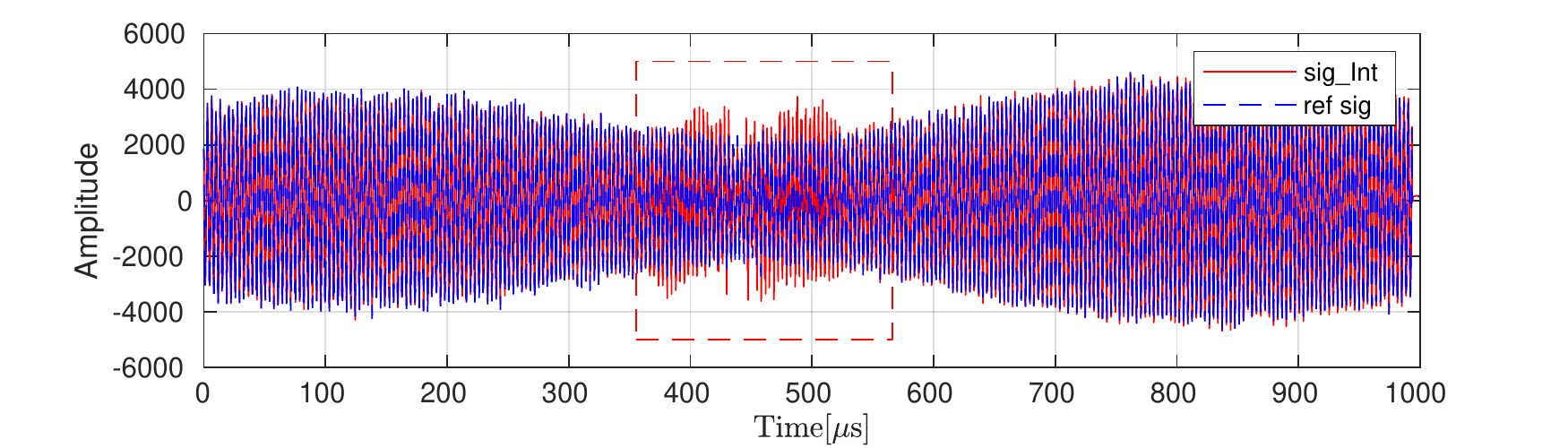}
		\label{fig:exp_Chimney_rawData}
	}
	
	\vspace{-3mm}
	\subfloat[]{
		\includegraphics[width=0.48\textwidth]{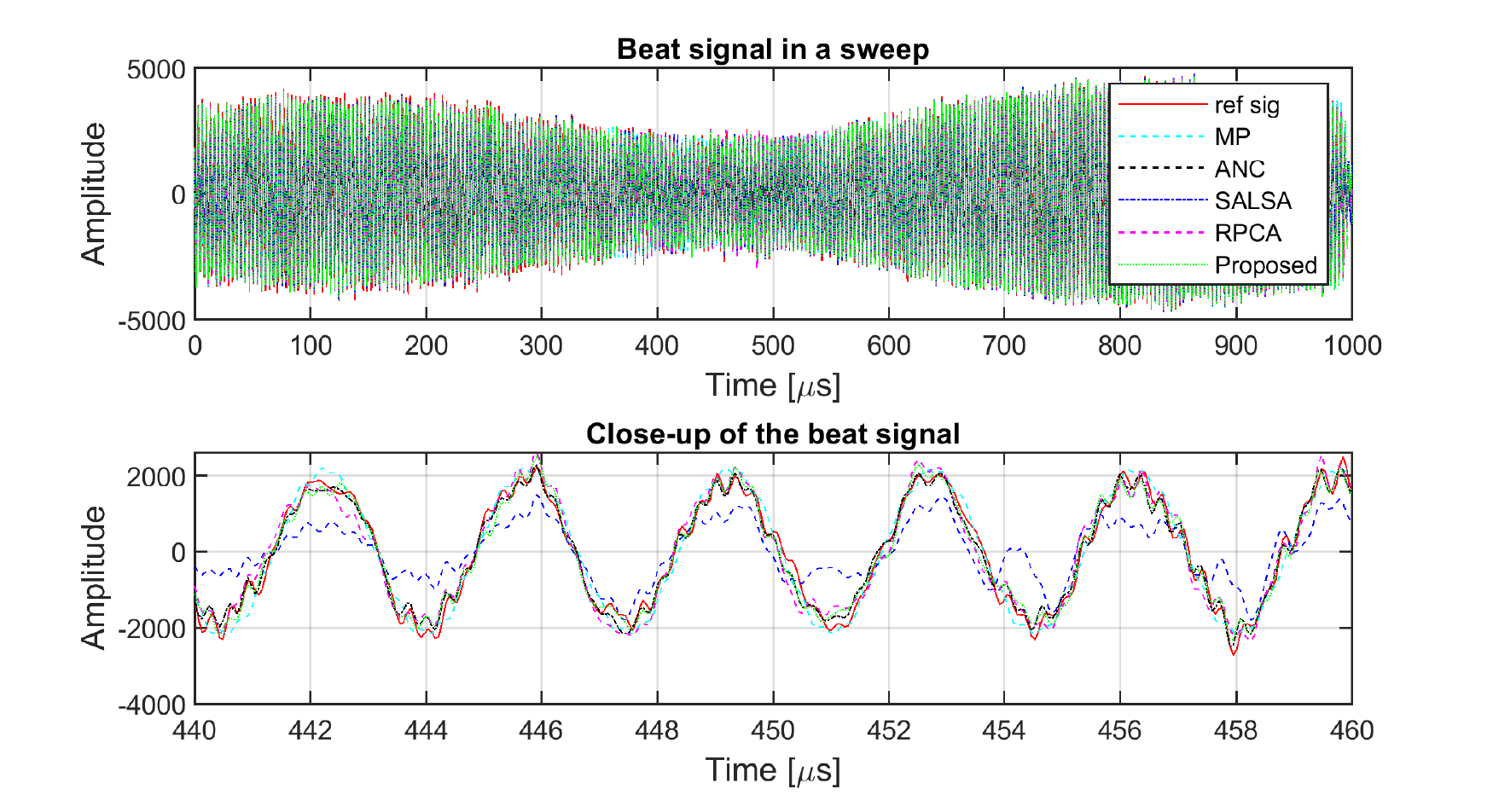}
		\label{fig:exp_Chimney_beatSig_AftMitig}
	}
	\caption{\protect\subref{fig:exp_Chimney_rawData} The beat signals with and without the interference, where the dashed-line rectangle indicates the region of  interference-contaminated samples. \protect\subref{fig:exp_Chimney_beatSig_AftMitig}
		The recovered beat signals of the chimney observation after interference mitigation. The upper panel shows the beat signals recovered by the MP-, ANC, SALSA-, RPCA-based methods and the proposed approach while the lower panel is the close-up of the beat signals in the interval of $[440,460]\mu s$ in the upper panel.}
	\label{fig:exp_Chimney_beatSig_IM}
\end{figure}

\begin{figure}[!t]
	\centering
	\vspace{-3mm}
	\subfloat[]{\hspace{-6mm}
		\includegraphics[width=0.56\textwidth]{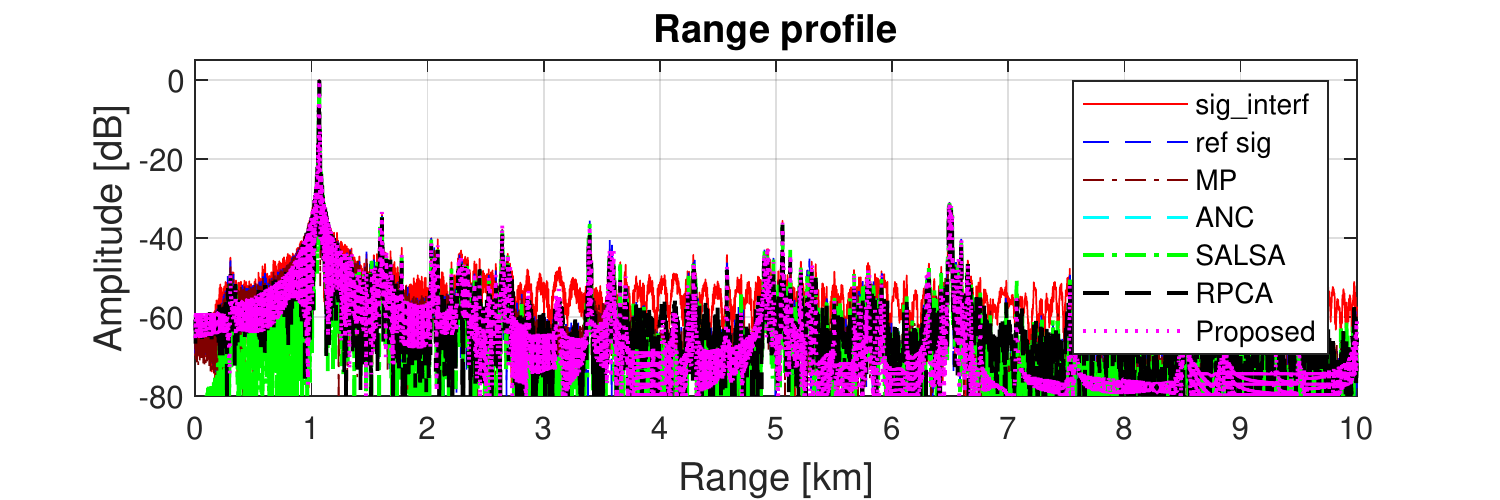}
		\label{fig:exp_Chimney_RangeProfile}
	}
	
	\vspace{-2mm}
	\subfloat[]{
		\includegraphics[width=0.4\textwidth]{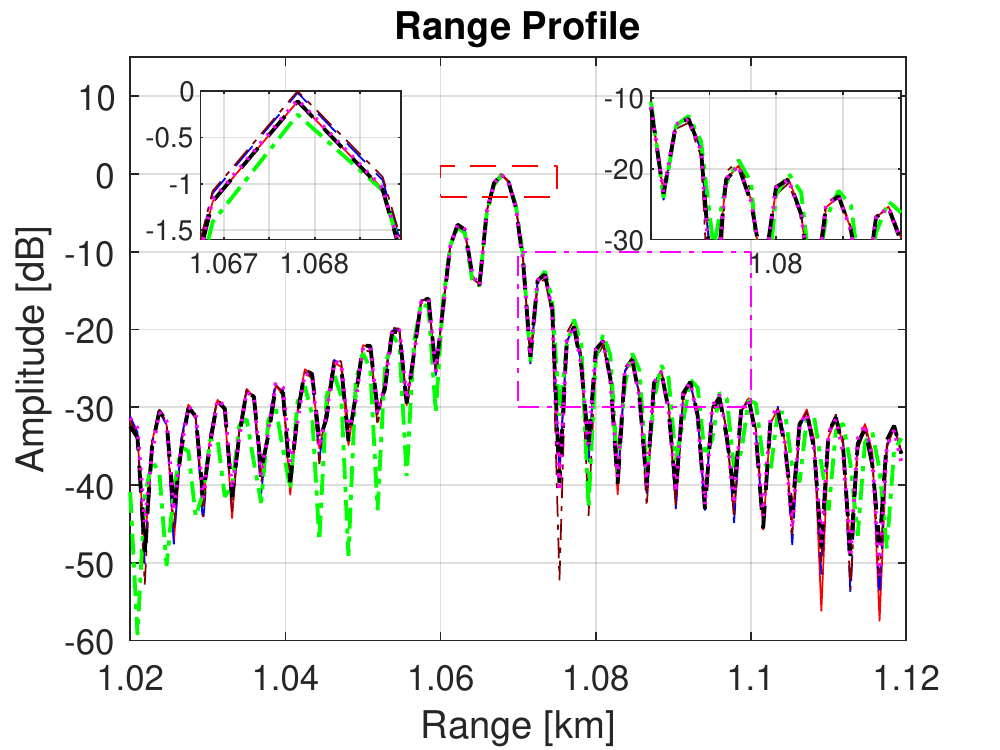}
		\label{fig:exp_Chimney_Range_profile_rec_1km}
	}
	
	\vspace{-2mm}
	\subfloat[]{
		\includegraphics[width=0.4\textwidth]{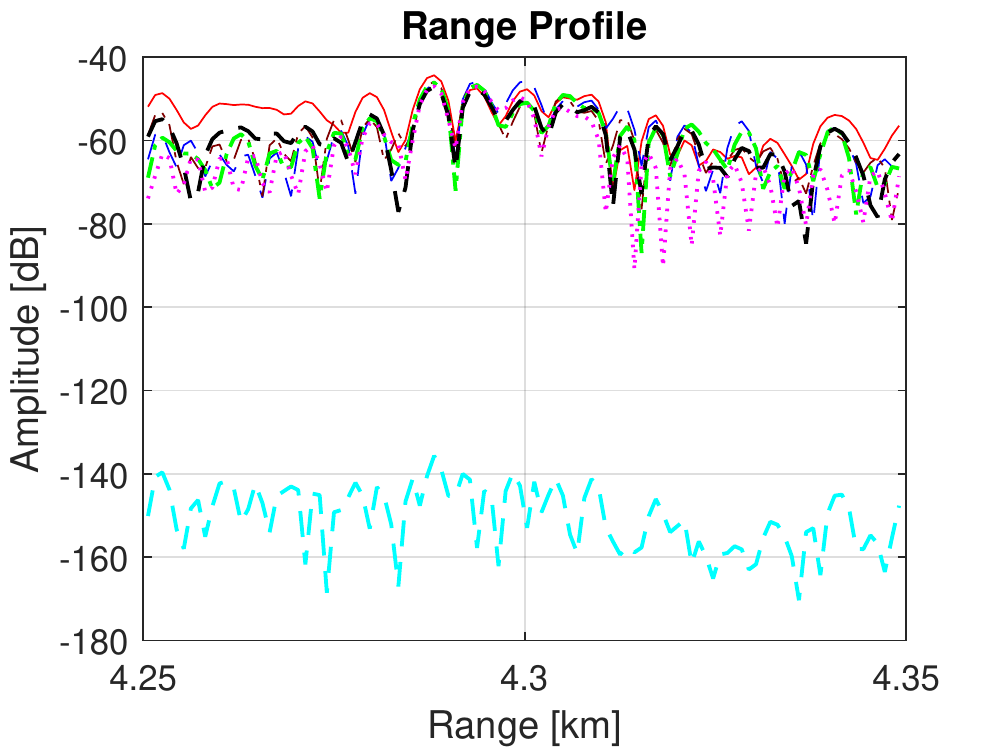}
		\label{fig:exp_Chimney_Range_profile_rec_4km}
	}
	\caption{\protect\subref{fig:exp_Chimney_RangeProfile} The range profiles of the chimney scenario obtained with the interference-contaminated beat signal, clean reference signal, and the recovered signals after interference mitigation with the ANC, MP-, SALSA-, RPCA-based methods, and the proposed approach. \protect\subref{fig:exp_Chimney_Range_profile_rec_1km} and \protect\subref{fig:exp_Chimney_Range_profile_rec_4km} show the zoomed-in view of the range profiles of targets at the distance of   $1.07\,\mathrm{km}$ and $4.3\,\mathrm{km}$ in \protect\subref{fig:exp_Chimney_RangeProfile}, respectively.}
	\label{fig:exp_Chimney_RP}
\end{figure}

\subsection{Experiment 1: Stationary isolated target (Chimney)} \label{sec:exp_1_chimney}
In this experiment, the PARSAX radar was set to observe a tall industrial chimney which is about $1.07\,\mathrm{km}$ away. The acquired HV-pol beat signal with the interference caused by the VV-pol component scattered from the illuminated scene is shown in Fig.~\ref{fig:exp_Chimney_beatSig_IM}\subref{fig:exp_Chimney_rawData}, where the interference-contaminated samples are indicated by a dashed-line rectangle. For comparison, the beat signal without interference was also collected by using only the H-pol transmitting channel to transmit up-chirp signal but turning off the V-pol transmitting channel in a consecutive sweep with a time delay of 1\,ms, which is presented as a reference (i.e., ``ref sig'' in Fig.~\ref{fig:exp_Chimney_beatSig_IM}\subref{fig:exp_Chimney_rawData}). Note as the experimental data were collected for a wild real scene where some dynamic objects may exist, the reference signal acquired after a time delay of 1\,ms may be, strictly speaking, not exactly the same as the useful signals due to the change of the scene. Nevertheless, for a slow-changing scene as in our case, the impact of the change of the scene in such a short time delay is almost negligible; thus, the acquired beat signal without interferences can be a ``valid'' reference.

Then, the interference-contaminated beat signal was processed with the ANC, SALSA-, RPCA-based methods and the proposed approach with empirically chosen optimal (hyper-) parameters. Specifically, the ANC used a 200-sample adaptive filter and the SALSA-based method utilized the same parameters as in section~\ref{sec:point_target_simu} except for a different regularization parameter $\lambda=0.5$. For the RPCA-based method, $\mu=0.0001$ was used for the data consistency constraint and $\delta=1\times10^{-6}$ for controlling the termination of the iterative algorithm. Meanwhile, for the proposed approach, almost the same parameters in section~\ref{sec:point_target_simu} were used except that $\tau$, $\beta$ and $\mu$ were initialized with 10, 0.125, and 0.0001, respectively.
After handling the interference mitigation by the ANC, SALSA-, RPCA-based methods and the proposed approach, the acquired beat signals are shown in Fig.~\ref{fig:exp_Chimney_beatSig_IM}\subref{fig:exp_Chimney_beatSig_AftMitig}. As only one interference appears in the acquired beat signal, the reconstructed beat signal with the matrix pencil (MP)-based method in \cite{JWang2020} is presented for comparison. Visually, the MP-, SALSA-, RPCA-based methods and the proposed approach all get satisfactory results which have very good agreement with the reference signal in contrast to that obtained with the ANC (see the lower panel in Fig.~\ref{fig:exp_Chimney_beatSig_IM}\subref{fig:exp_Chimney_beatSig_AftMitig}). Quantitatively, the SINRs of the recovered beat signals with the ANC, MP-, SALSA-, RPCA-based methods and the proposed approach are $2.93\,\mathrm{dB}$, $21.40\,\mathrm{dB}$, $17.87\,\mathrm{dB}$, $20.77\,\mathrm{dB}$ and $20.78\,\mathrm{dB}$, respectively. The corresponding correlation coefficients $|\rho|$ are $0.7008$, $0.9964$, $0.9919$, $0.9958$, and $0.9958$. So according to the values of the SINR and $|\rho|$, the MP-based method gets the most accurate result, which is slightly better than that acquired with the RPCA-based method and the proposed one. However, the MP-based method requires to first detect the location of the interference in a sweep, and then cut out the interference-contaminated signal samples and reconstruct them based on the rest. Although we assume the location of the interference is accurately detected in this experiment, it is generally a challenge to precisely distinguish the interference from the useful signal. By contrast, the RPCA-based method and the proposed approach have no such requirement, which can tackle the interference mitigation blindly.      

To further illustrate the accuracy of the recovered beat signals, the corresponding range profiles of targets are obtained by taking the FFT of them with respect to time, as shown in Fig.~\ref{fig:exp_Chimney_RP}\subref{fig:exp_Chimney_RangeProfile}. Compared to the range profile of the reference signal, both the MP-based method and the proposed approach significantly suppress the interference and reduce the ``noise'' floor (Fig.~\ref{fig:exp_Chimney_RP}\subref{fig:exp_Chimney_RangeProfile}). Their resultant range profiles have a very good agreement with the reference one (see the zoomed-in range profiles of the chimney and the weak scattering object in Fig.~\ref{fig:exp_Chimney_RP}\subref{fig:exp_Chimney_Range_profile_rec_1km} and \subref{fig:exp_Chimney_Range_profile_rec_4km}, respectively). Although both the SALSA- and RPCA-based methods effectively mitigate the interference, the range profiles produced by their recovered signals have slightly higher ``noise'' floor compared to that obtained with the MP-based method and the proposed approach (Fig.~\ref{fig:exp_Chimney_RP}\subref{fig:exp_Chimney_RangeProfile}). Moreover, the signal acquired by handling the interference mitigation with the SALSA-based method leads to small (i.e., about $0.2\,\mathrm{dB}$) loss of the peak power of the range profile of the chimney (Fig.~\ref{fig:exp_Chimney_RP}\subref{fig:exp_Chimney_Range_profile_rec_1km}) while the ANC completely suppresses the signal of the weak scattering object at the distance of $4.3\,\mathrm{km}$ (Fig.~\ref{fig:exp_Chimney_RP}\subref{fig:exp_Chimney_Range_profile_rec_4km}).

\subsection{Experiment 2: Distributed target (Rain)}  
A rain storm was observed by steering the PARSAX radar pointing to the zenith. The fully polarimetric data were acquired by simultaneously transmitting and receiving both horizontally and vertically polarized signals. Here we use the same HV-polarimetric data in \cite{JWang2020} to demonstrate the interference mitigation, which are contaminated by the strong VV-polarimetric signals arriving at the receiver at the same time. This rain dataset contains radar signals measured in one coherent processing interval, i.e., 512 FMCW sweeps.
As the rain droplets are moving targets, the range Doppler (R-D) map is generally used to characterize their distribution. Considering the relatively stable interference location in each sweeps and avoiding possible negative effects of the inaccuracy of the interference suppression, the processing flow we take is: first take the FFT of the HV signals along the slow time and then handling interference mitigation of the signals in each Doppler frequency bin followed by an FFT operation along the fast time to get the focused R-D map  \cite{JWang2020}.

\begin{figure}[!t]
	\centering
	\includegraphics[width=0.42\textwidth]{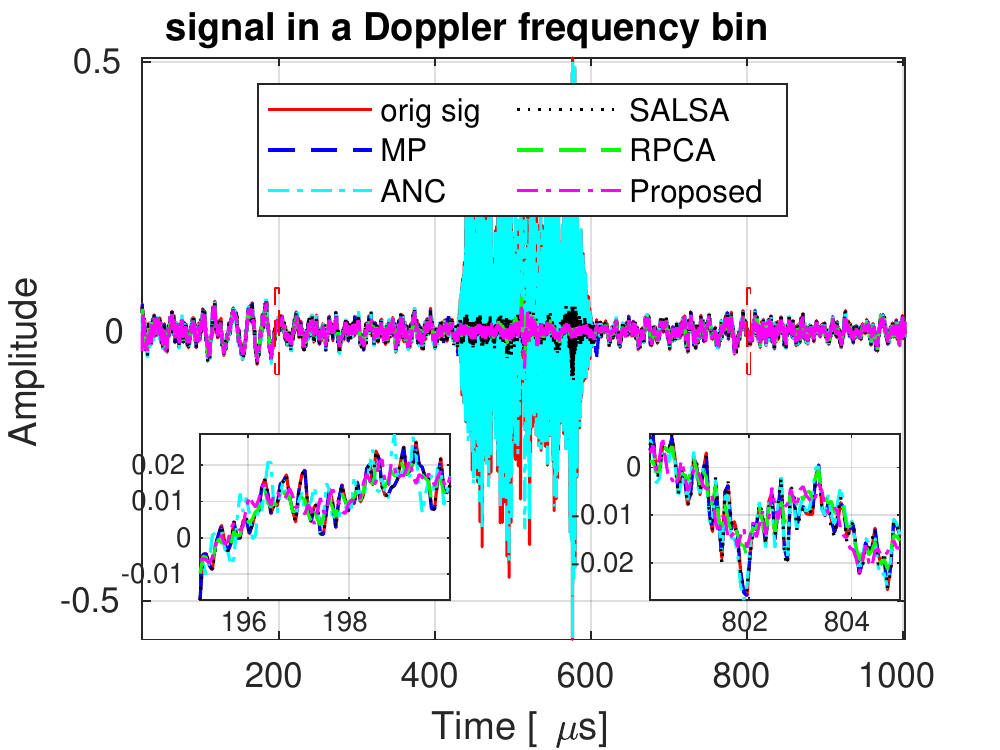}
	\caption{Signal in a Doppler frequency bin. Both the signals before  (i.e., orig sig) and after interference mitigation are presented.}
	\label{fig:exp_Rain_sig_DopplerBin}
\end{figure}

To perform interference mitigation for signals in each Doppler bin, the ANC was tuned to utilize a 50-sample adaptive filter. For the SALSA-based method, the regularization parameters $\lambda$ and $\mu$ took the values of 0.875 and 1, respectively; meanwhile, the STFT in its each iteration used a 256-sample rectangular window with 243 samples (i.e., 95\%) of overlap between adjacent segments and 256-point FFT. Moreover, the RPCA was implemented by using the same parameters as in section~\ref{sec:exp_1_chimney} except for $\mu=0.005$. For the proposed approach, $\beta=1$, $\mu=8$, $\tau=0.08$ and $\delta=1\times10^{-6}$ were initialized. To speed up the convergence of the algorithm, $\mu$ and $\beta$ were gradually increased with the parameters $k_\mu=1.1$, $k_\beta=1.8$ and $L=10$.
Fig.~\ref{fig:exp_Rain_sig_DopplerBin} shows the raw beat signal in a Doppler frequency bin and the recovered counterparts after dealing with interference mitigation with the ANC, SALSA-, RPCA-based methods and the proposed approach. Similar to Experiment 1, the result of the MP-based method in \cite{JWang2020} is also presented for comparison. One can see that the ANC almost fails to suppress the interference in this experiment while all the other approaches significantly mitigate the interference. As the MP-based method only cuts out the interference-contaminated measurements and then reconstructs the samples of useful signals in the cut-out region, the rest of signal samples keeps the same as the original signal (see the insets in Fig.~\ref{fig:exp_Rain_sig_DopplerBin}). By contrast, the SALSA-, RPCA-based methods and the proposed approach tackle the interference mitigation as a signal separation problem, and both the interference and noise could be separated and removed from the useful signals. Thus, the recovered signal samples in the interference-free region could slightly deviate from the original ones due to the de-noising effect. As the ground truth signal is unavailable, it is difficult to directly evaluate the accuracy of these recovered signals. 

\begin{figure}[!t]
	\centering
	\subfloat[]{ \hspace{-3mm}
		\includegraphics[width=0.25\textwidth]{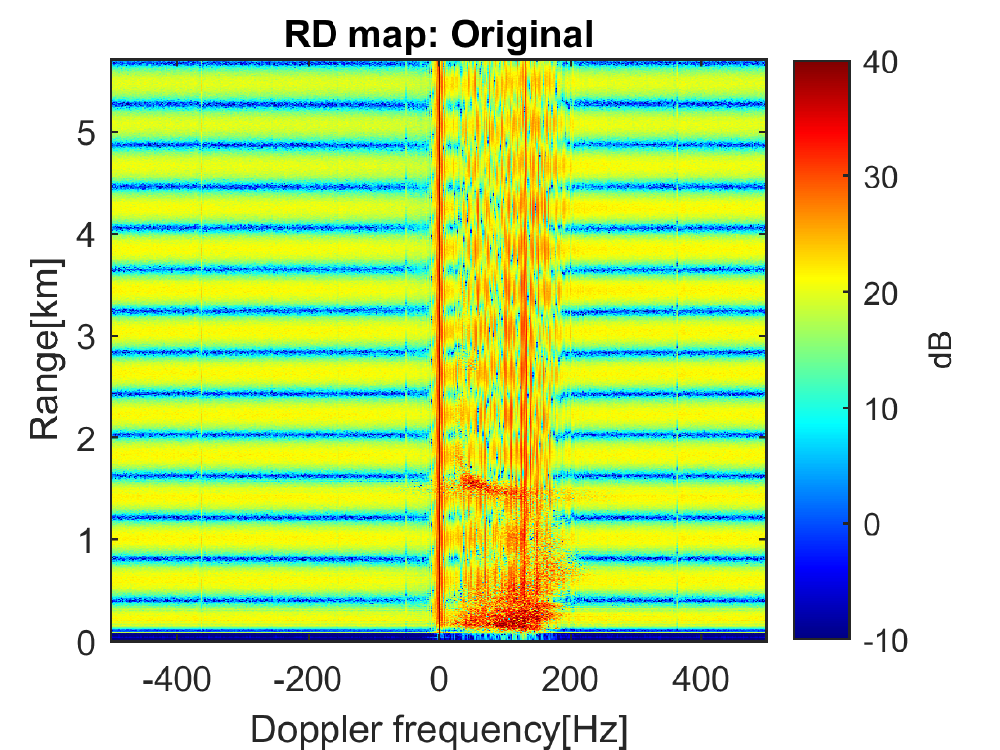}
		\label{fig:exp_rain_RD_Origin_HV}}
	\subfloat[]{\hspace{-4mm}
		\includegraphics[width=0.25\textwidth]{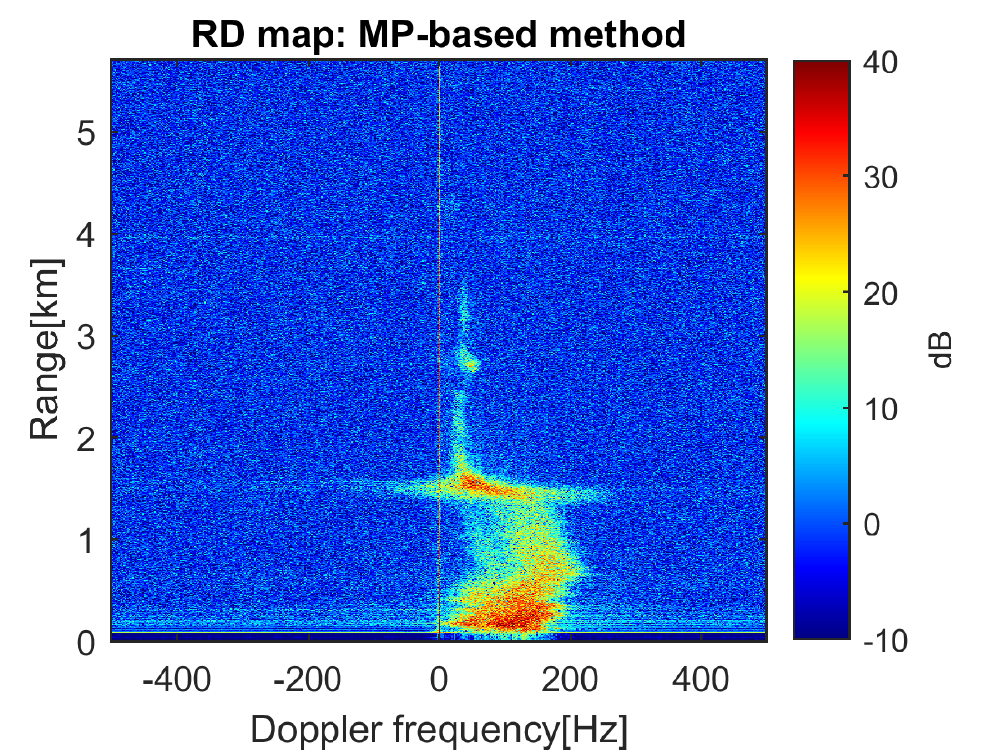}
		\label{fig:exp_rain_RD_MP_HV}}

	\subfloat[]{ \hspace{-3mm}
		\includegraphics[width=0.25\textwidth]{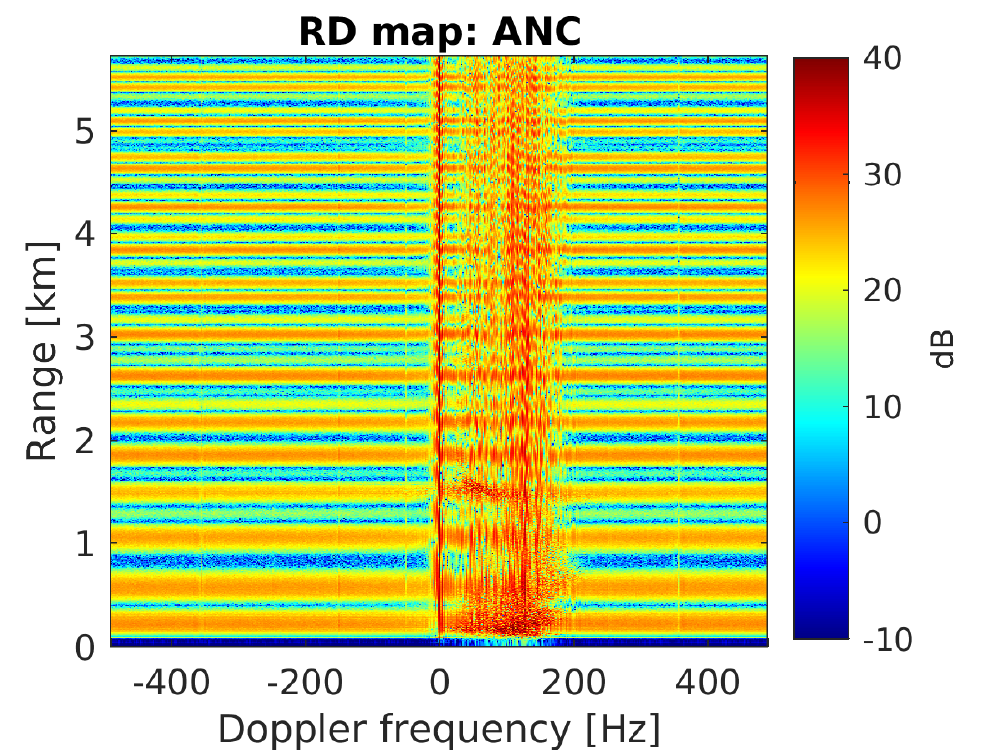}
		\label{fig:exp_rain_RD_ANC}
	}
	\subfloat[]{ \hspace{-4mm}
		\includegraphics[width=0.25\textwidth]{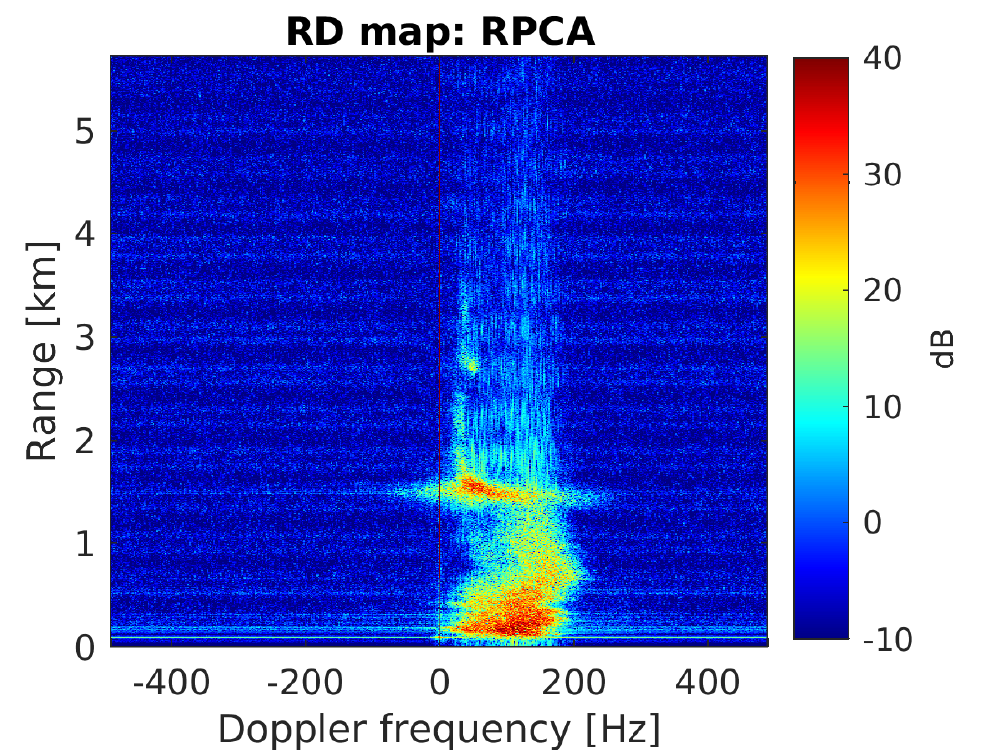}
		\label{fig:exp_rain_RD_RPCA}
	}

	
	\subfloat[]{ \hspace{-3mm}
		\includegraphics[width=0.25\textwidth
		]{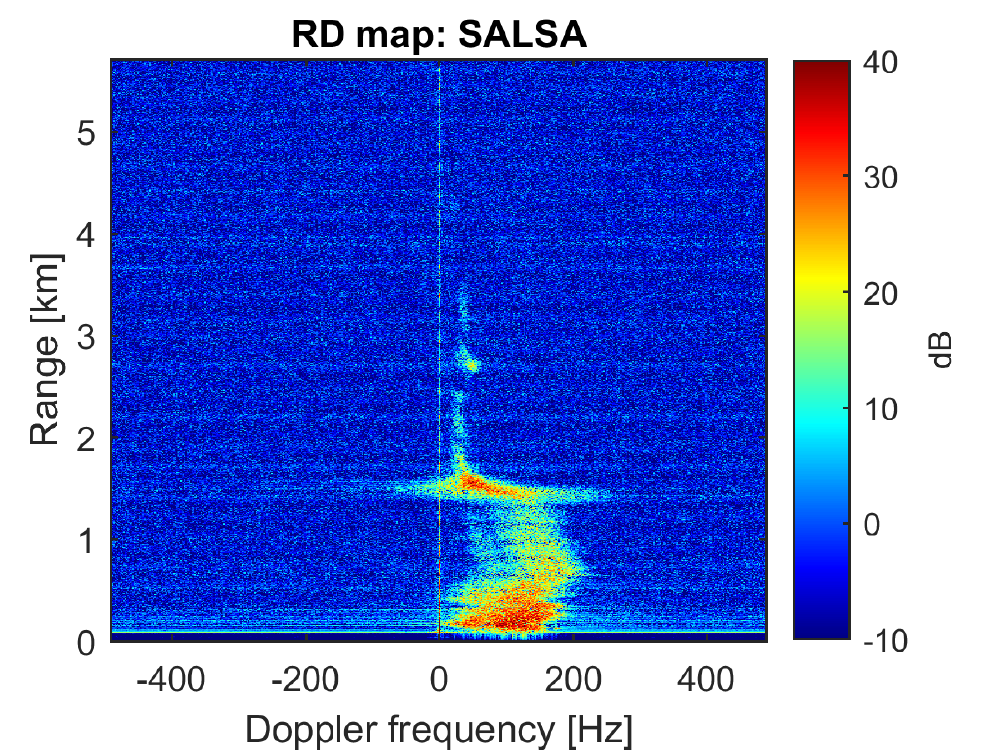}
		\label{fig:exp_rain_RD_salsa}
	}
	\subfloat[]{ \hspace{-4mm}
		\includegraphics[width=0.25\textwidth]{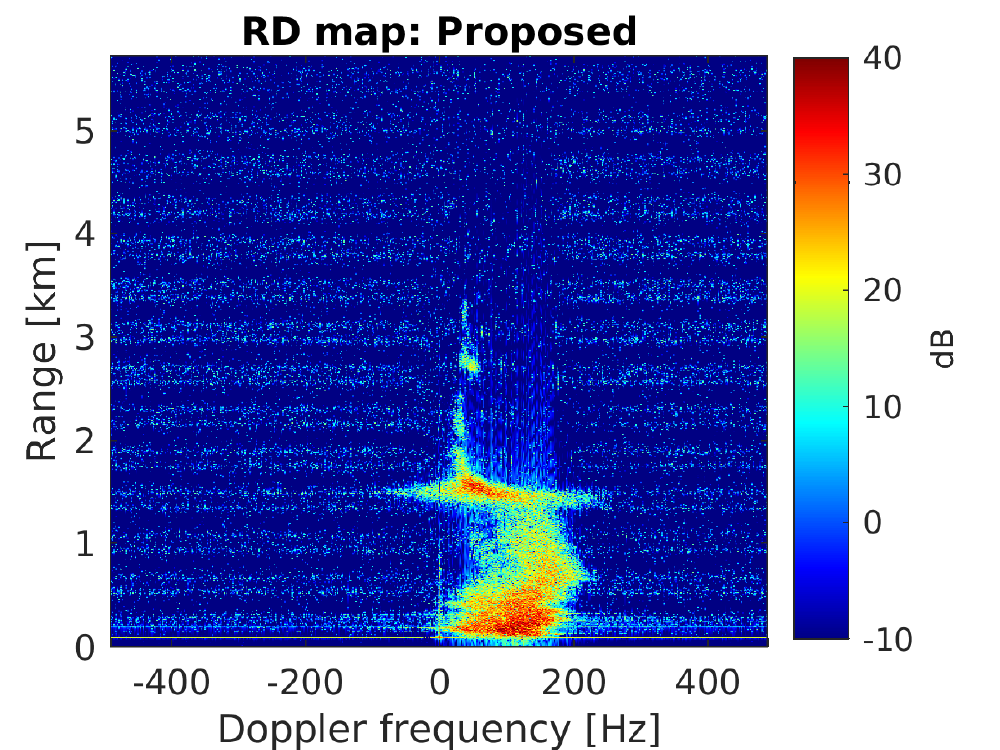}
		\label{fig:exp_rain_RD_MC}
	}
	
	\caption{Range-Doppler map of the rain. \protect\subref{fig:exp_rain_RD_Origin_HV} is obtained with the raw data, \protect\subref{fig:exp_rain_RD_MP_HV} is formed with the signals recovered by the MP-based method, \protect\subref{fig:exp_rain_RD_ANC} ANC, \protect\subref{fig:exp_rain_RD_RPCA}the RPCA-based method, \protect\subref{fig:exp_rain_RD_salsa} the SALSA-based method, and \protect\subref{fig:exp_rain_RD_MC} the proposed approach. }
	\label{fig:exp_rain_Range-Doppler}
\end{figure}

Nevertheless, with the signals recovered by each method, the R-D maps can be reconstructed and compared to assess their performance of interference mitigation. Fig.~\ref{fig:exp_rain_Range-Doppler} shows the focused R-D maps with the recovered signals as well as the original ones. It is clear that the R-D map of the rain droplets is still completely masked by the strong interference (see Fig.~\ref{fig:exp_rain_Range-Doppler}\subref{fig:exp_rain_RD_ANC}). Comparing Fig.~\ref{fig:exp_rain_Range-Doppler}\subref{fig:exp_rain_RD_MP_HV}, \subref{fig:exp_rain_RD_RPCA}, \subref{fig:exp_rain_RD_salsa} and \subref{fig:exp_rain_RD_MC}, one can see that the MP-, SALSA-based method and the proposed approach get cleaner range-Doppler distribution of rain droplets than the RPCA-based method, especially in the area above the range of $2\,\mathrm{km}$ with Doppler frequency larger than $50\,\mathrm{Hz}$. Meanwhile, the R-D map obtained with the MP-based method has the most uniform background than that of the RPCA-, SALSA-based methods and the proposed approach. Although some remaining weak interference streaks can be observed in Fig.~\ref{fig:exp_rain_Range-Doppler}\subref{fig:exp_rain_RD_MC}, the proposed approach leads to lower background noise floor compared to that in Fig.~\ref{fig:exp_rain_Range-Doppler}\subref{fig:exp_rain_RD_MP_HV}. This is attributed to the denoising effect of the proposed approach  during the signal separation. In principle, the remaining weak interference can be further suppressed by searching the optimal regularization parameters $\mu$, $\beta$ and $\tau$. In addition, the proposed approach effectively suppresses the zero-Doppler interference compared to the MP-, SALSA-, and the RPCA-based methods.

\section{Conclusion}  \label{sec:conclusion}

In this paper, an approach based on sparse and low-rank decomposition of the Hankel matrix is proposed for interference mitigation of FMCW radars.  Compared with the existing interference nulling and reconstruction methods, the proposed approach does not need to detect the location of interferences and is able to blindly handle complex interference mitigation problem with multiple interferers. In contrast to the FFT-based signal separation methods (for instance, the SALSA-based method) which exploit the sparsity of the signals and the interference on the regular discrete grids of Fourier bases, the proposed algorithm utilizes the gridless optimization. So it improves the accuracy of the recovered signal by avoiding the possible mismatch between the spectrum of the signal and the discrete grid. Moreover, the numerical simulations demonstrate that  the proposed approach can substantially suppress the interferences with the duration up to $50\%$ of the signal sweep based on the time sparsity assumption. Note that the proposed approach is readily to be extended to exploit the sparsity of the interference in a transformed domain by replacing the regularization term $|\mathbf{i}|_1$ with the corresponding transformed counterpart $|\mathcal{T}(\mathbf{i})|_1$. Finally, it should be mentioned that the three regularization parameters involved in the proposed approach affect both the convergence rate and its interference mitigation performance, which are usually empirically selected in practice. However, choosing the optimal values for them may be not trivial for practical applications. So how to automatically select the optimal values of these regularization parameters would be a further research topic in future.


%





\ifCLASSOPTIONcaptionsoff
  \newpage
\fi




\bibliographystyle{./hieeetr.bst}

%
\bibliography{Reference}

%








\end{document}